\documentclass[aps,superscriptaddress,showpacs,nofootinbib,eqsecnum,prd,notitlepage]{revtex4}

\usepackage{amssymb}
\usepackage{amsmath}
\usepackage{natbib}
\usepackage{epsfig}
\usepackage{graphicx}
\usepackage{dcolumn}
\usepackage{bm}
\usepackage[english]{babel}
\usepackage[latin1]{inputenc}
\usepackage{eucal}
\usepackage{hyperref}
\usepackage{verbatim}
\usepackage{latexsym}
\usepackage{xcolor}
\usepackage{color}

\begin{document}

\title{BEC-BCS crossover in a cold and magnetized two color NJL model}

\author{Dyana C. Duarte}
\affiliation{ Departamento de F\'isica, Universidade Federal de Santa Maria, 97105-900,
Santa Maria, RS, Brazil}

\author{P. G. Allen}
\affiliation{Department of Theoretical Physics, Comisi\'on Nacional de Energ\'ia At\'omica,
Av. Libertador 8250, 1429 Buenos Aires, Argentina}

\author{R. L. S. Farias}
\affiliation{ Departamento de F\'isica, Universidade Federal de Santa Maria, 97105-900,
Santa Maria, RS, Brazil}
\affiliation{Department of Physics, Kent State University, Kent, OH 44242, United States}

\author{Pedro H. A. Manso}
\affiliation{Departamento de F\'{\i}sica Te\'orica, Universidade do
  Estado do Rio de Janeiro, 20550-013 Rio de Janeiro, RJ, Brazil}
\affiliation{Centro Federal de Educa\c{c}\~ao Tecnol\'ogica Celso Suckow da Fonseca,
Campus Maria da Gra\c{c}a,Rua Miguel \^{A}ngelo 96, 20785-223 Rio de Janeiro, RJ,
Brazil}

\author{Rudnei O.  Ramos}
\affiliation{Departamento de F\'{\i}sica Te\'orica, Universidade do
  Estado do Rio de Janeiro, 20550-013 Rio de Janeiro, RJ, Brazil}

\author{N. N. Scoccola}
\affiliation{Department of Theoretical Physics, Comisi\'on Nacional de Energ\'ia At\'omica,
Av. Libertador 8250, 1429 Buenos Aires, Argentina}
\affiliation{Department of Theoretical Physics, Comisi\'on Nacional de Energ\'ia At\'omica,
\affiliation{Av. Libertador 8250, 1429 Buenos Aires, Argentina}
CONICET, Rivadavia 1917, 1033 Buenos Aires, Argentina}
\affiliation{Universidad Favaloro, Sol\'is 453, 1078 Buenos Aires, Argentina}

\begin{abstract}

The BEC-BCS crossover for a  NJL model with diquark interactions
is studied in the presence of an external magnetic field.
Particular attention is paid to different regularization schemes
used in the literature.  A thorough comparison of results is
performed for the case of a cold and magnetized  two-color NJL
model. According to our results, the critical chemical potential
for the BEC transition exhibits a clear inverse magnetic catalysis
effect for magnetic fields in the range $ 1 \lesssim  eB/m_\pi^2
\lesssim 20 $. As for the BEC-BCS crossover, the corresponding
critical chemical potential is very weakly sensitive to magnetic
fields up to $eB \sim 9\ m_\pi^2$, showing a much smaller inverse
magnetic catalysis as compared to the BEC transition, and displays
a strong magnetic catalysis from this point on.

\end{abstract}


\maketitle

\section{Introduction}

Although a considerable amount of theoretical and experimental
work has been devoted to the subject, the phase diagram of Quantum
Chromodynamics (QCD) still remains poorly understood. {}From the
theoretical point of view, one of the main reasons for this state
of affairs is that the ab-initio lattice QCD approach has
difficulties to deal with the region of medium/low temperatures
and moderately high densities, owing to the so-called ``sign
problem''~\cite{Karsch:2001cy,Muroya:2003qs}. Thus, most of the
present knowledge about the strongly interacting matter phase
diagram arises from the study of effective models~\cite{review}.
This is because effective models offer the possibility of obtaining
predictions for the
transition features at regions that are not accessible through
lattice techniques. In this context, in the last years several
works have considered that, at low temperatures, the transition
between the chirally broken phase at low densities and the color
superconducting phase at large densities proceeds in a smooth way
instead of being a strong first-order transition, as is more
commonly believed. Interestingly, this opens up the possibility
that, as density increases, quark matter undergoes a crossover
between a regime where diquark pairs form difermion molecules in
Bose-Einstein condensation (BEC) and a weakly coupled
Bardeen-Cooper-Schrieffer (BCS) superfluid
regime~\cite{Nishida:2005ds}.  One of the models where this kind
of phenomena has been more actively investigated is the well-known
Nambu-Jona-Lasinio (NJL) model~\cite{reports}, where
gluon-mediated interactions are replaced by effective local
quark-quark interactions.

The actual existence of a BEC-BCS crossover in the three-color
($N_c=3$) NJL
model requires, however, a quite strong quark pairing interaction.
Let us recall that the application of the {}Fierz transformations
to the effective one gluon exchange interaction (OGE) leads to $r
\equiv G_D/G_S = 3/4$, where $G_S$ and $G_D$ are the scalar
quark-antiquark and the diquark coupling constants, respectively.
This value is usually taken as a reference in most model
calculations~\cite{Buballa:2003qv}. On the other hand, the
existence of a BEC-BCS crossover at finite baryon chemical
potential requires $r \gtrsim
1$~\cite{Sun:2007fc,Kitazawa:2007zs}. It has been argued, however,
that in the flavor SU(3) NJL model the axial anomaly might induce
a BEC-BCS crossover for more conventional values of
$r$~\cite{Abuki:2010jq,Basler:2010xy}. Other possibility has been
proposed in \cite{Ferrer:2014ywa}.  Contrary to what happens for
$N_c=3$, the situation for the $N_c=2$ NJL model is quite clear.
In this case, the {}Fierz transformation of the OGE interaction
leads to $r=1$, as shown in Ref.~\cite{Ratti:2004ra}. {}For this
value of $r$, the authors in Ref.~\cite{Sun:2007fc} find that as
the quark chemical potential $\mu$ increases, there is first, at
$\mu=m_\pi/2$, a second-order transition from the chirally broken
phase to the BEC phase, followed by the BEC-BCS crossover at a
somewhat larger chemical potential.  This result was confirmed in
Refs.~\cite{Brauner:2009gu,He:2010nb}.

Despite the fact the two-color NJL model might only share some
qualitative similarities with real three-color QCD, it is still
a valuable model for studying in general. In fact, studies with two-color
QCD like models, in particular in the context of the NJL model, have
been quite popular (for a recent review, see, e.g., Ref.~\cite{lhe_review}
and references therein). Because of the different gauge group
of two-color QCD (which has only pseudoreal, or real representations)
as compared to the three-color case, the fermion
determinant in the former remains real for nonvanishing chemical potentials
(for baryons or quarks). Hence, the model does not suffer from the
fermion-sign problem that plagues the three-color QCD and its phase diagram can be studied through
lattice Monte-Carlo simulations~\cite{Kogut:2001na,Buividovich:2008wf,Ilgenfritz:2012fw}.
Besides, the two-color NJL model can be seen as a relativistic analogue
of the low energy non-relativistic BCS-BEC crossover in condensed
matter fermionic systems. Given the above properties, we hope to learn
some of the related physics associated with diquarks condensation in real
QCD by using this simpler model and the generic properties connected
to a BEC-BCS transition in fermionic systems in general.

In the present work, we will be mainly interested in the role played by a
magnetic field in the BEC-BCS transition problem.  This is mostly
motivated by the realization that strong magnetic fields may be
produced in several physically relevant situations. {}For example,
the compact stellar objects believed to be the source of intense
$\gamma$ and X-rays, the magnetars, are expected to bear fields of
the order of $10^{13}-10^{15}$  G at their surface\cite{Duncan:1992hi}, reaching
values several orders of magnitude greater at their
center~\cite{Shapiro:1991hi,Bandyopadhyay:1997kh,Ferrer:2010wz}.
Moreover, such strong magnetic
fields are also expected in present and future experiments at the
RHIC, NICA and FAIR facilities, designed to probe the phase
diagram of strongly interacting matter at low temperatures and
intermediate-to-large densities.

So far, the effect of magnetic field on the BEC/BCS crossover has
been analyzed in the context of a two-channel
model~\cite{Wang:2010uj}, where fermion and boson degrees of
freedom are introduced from the beginning in the corresponding
effective Lagrangian density. The aim of the present study is to
go beyond this by considering a single channel model. Since this
is already a quite complicated issue by itself, we will work in
the framework of the two-color NJL model,  where the existence of
a BEC-BCS crossover for conventional model parameterizations is
firmly established.  We recall here that, in contrast to what
happens in $N_c=3$ QCD, this theory allows for lattice simulations
at finite chemical potential.  Results from this model regarding
the diquark condensation in the absence of magnetic fields have
been compared, e.g. in Ref.~\cite{Ratti:2004ra}, with lattice
simulations of $N_c=2$ QCD~\cite{Kogut:2001na}. It turns out that
the predictions of the NJL model for the diquark condensate are
similar than those of chiral perturbation theory~\cite{ChPT}, as
far the agreement with lattice results are concerned. The NJL
model is therefore interesting to investigate the general symmetry
properties associated with two-color quark matter and their role
in the phase transitions. Moreover, some aspects of the impact of
the presence of external magnetic fields on the properties of
two-color quark matter at finite temperature and vanishing
chemical potential have been investigated using both effective
models~\cite{Andersen:2012jf} and lattice
simulations~\cite{Buividovich:2008wf,Ilgenfritz:2012fw}.

This paper is organized as follows. In Sec.~\ref{Thpot} we
introduce a two-color and two-flavor NJL-type model in the
presence of a magnetic field and the corresponding thermodynamic
potential to study the phase structure and, in
particular, the BEC-BCS transition. Some possible regularization
schemes used in the literature are also given.  In
Sec.~\ref{results} we give the results of our analysis and also
make a comparison between different regularizations used when regarding the
critical quantities and the crossover. In Sec.~\ref{results2},
we interpret the results obtained and check for their consistency.
Our concluding remarks are
given in Sec.~\ref{concl}.  One appendix is included, where some
of the technical details are given.

\section{The $N_c=2$ NJL model in the presence of an external magnetic field}
\label{Thpot}

We start by specifying the model and its parameters.  We consider
a two-color and two-flavor NJL-type model that includes
scalar-pseudoscalar and color pairing interactions. The
corresponding Lagrangian density, in the presence of an external
electromagnetic field,  is given by

\begin{eqnarray}
 \mathcal{L} & = & \bar{\psi}\left(i\ {\rlap/\!D} -m_c\right)\psi + G_S
 \left[\left(\bar{\psi}\psi\right)^2  +
   \left(\bar{\psi}i\gamma_5\tau\psi\right)^2\right]
+ G_D \left(\bar{\psi}i\gamma_5 \tau_2 t_2 C\bar{\psi}^T\right)
 \left(\psi^T Ci\gamma_5\tau_2 t_2 \psi\right)~.
 \label{lag1}
\end{eqnarray}
Here, $C = i\gamma_0\gamma_2$ is the charge conjugation matrix and
$m_c$ is the current fermion mass, which we take to be equal for
both flavors. Moreover, $\tau_i$ and $t_i$ are the Pauli matrices
in flavor and color spaces, respectively. As usual, we will assume that the
interaction terms are derived from an effective local coupling
between color current terms. The two coupling
constants $G_S$ and $G_D$, for the
present two-color NJL model~\cite{Ratti:2004ra} can be related simply as  $G_S = G_D = G$.
This relation has to be imposed in order to have the
symmetries of two-color QCD, and must be satisfied regardless of
the type of the effective interaction that we start from. The coupling of
the quarks to the electromagnetic field ${\cal A}_\mu$ is
implemented through the covariant derivative,
$D_{\mu}=\partial_\mu - i  Q {\cal A}_{\mu}$, where $Q$ is the
usual quark charge matrix $Q =\mbox{diag}(q_u,q_d)$. In the
present $N_c=2$ model we choose the charges to be $q_u = e$ and
$q_d = -e$. In this way, as in Ref.~\cite{Wang:2010uj}, the charged fermions can be
considered to mimic the rotated charged quarks that pair to form
the electrically neutral Cooper pairs in actual QCD. Note,
however, that while in the latter case quark matter will still
behave as a color superconductor, two-color Cooper pairs are
color-singlets. Therefore, in the two-color QCD we are dealing
with a true superfluid where just a global (baryon number)
symmetry is spontaneously broken. {}Finally, we consider that the system is in the presence of a
static and constant magnetic field in the third-direction, namely,
we take ${\cal A}_\mu=\delta_{\mu 2} x_1 B$.

\subsection{The thermodynamical potential in the mean field approximation}

At vanishing magnetic field, the mean-field thermodynamic potential at
temperature $T$ and chemical potential $\mu$ is given
by~\cite{Sun:2007fc}

\begin{eqnarray}
 \Omega & = &
\Omega_0  - 8 T \sum_{s=\pm 1} \int\frac{d^3k}{(2\pi)^3}
\ln\left\{1 + \exp\left[-\frac{\sqrt{(E_k + s \ \mu)^2 + \Delta^2}}{T}\right]\right\}~,
 \label{potential}
\end{eqnarray}
where $\Omega_0$ is the zero temperature contribution,

\begin{equation}
 \Omega_0 = \frac{(m - m_c)^2 + \Delta^2}{4G} - 4 \sum_{s=\pm
   1}\int\frac{d^3k}{(2\pi)^3}\ \sqrt{(E_k + s \ \mu)^2 + \Delta^2}~,
 \label{Omega0}
\end{equation}
and $E_k = \sqrt{k^2 + m^2}$, with $m$ being the dressed quark mass.

The addition  of a finite constant
magnetic field can be easily implemented in the above expressions by considering the
following substitutions~\cite{Menezes:2008qt}

\begin{eqnarray}
&& 2 \int \frac{d^{3}k}{\left( 2\pi \right) ^{3}} \rightarrow \sum_{f
    = u}^{d}\frac{\left| q_{f}\right| B}{4\pi} \sum_{l = 0}^{\infty }
  \alpha_{l} \int_{-\infty}^{+\infty} \frac{dk_{3}}{ 2\pi }, \nonumber
  \\ && E_k \to E_{k_3,l} = \sqrt{k_3^2 + 2l|q_f|B + m^2}\,,
\label{Btrick}
\end{eqnarray}
where $\alpha_{l} = 2 - \delta_{l,0}$ takes into account the
degeneracy of the Landau levels.  Therefore, the resulting
zero-temperature mean-field thermodynamic potential in the presence of
a constant magnetic field reads

\begin{equation}
\Omega_{0}(m,\Delta,B,\mu) = \frac{\left(m - m_{c}\right)^{2} +
  \Delta^{2}}{4G} - 2 \sum_{f =
  u}^{d}\frac{\left|q_{f}\right|B}{4\pi} \sum_{s=\pm 1} \sum_{l =
  0}^{\infty}\alpha_{l} \int_{-\infty}^{+\infty}\frac{dk_{3}}{2\pi}
\sqrt{(E_{k_3,l} + s\ \mu)^2 + \Delta^2}~.
 \label{Omega0B}
\end{equation}

In the following, we will discuss the different ways to regularize the
thermodynamic potential when including the effects of an external magnetic field.

\subsection{Model parameters and regularization schemes}

Since the model under consideration is non-renormalizable, a
proper regularization scheme is required to avoid ultraviolet
divergences. As it will be discussed below for the regularization
procedures to be used in this work, this implies the existence of
a cutoff parameter $\Lambda$. Thus, the coupling constant
$G = G_S = G_D$, the current quark mass $m_c$ and $\Lambda$ form
a set of three parameters that must be specified in order to
proceed with the numerical calculations. In the case of the
$N_c=3$ NJL model analysis, these parameters are usually fixed such
as to reproduce the empirical values in vacuum for the pion mass
$m_\pi$, the pion decay constant $f_\pi$ and for the quark
condensate $\langle\bar qq\rangle_0$. The situation for the
$N_c=2$ case is not, however, so clear. Here, we follow the
procedure proposed in Ref.~\cite{Brauner:2009gu}, which is based
on the $N_c$ scaling of physical quantities. Using the fact that
$f_\pi$ is proportional to $\sqrt{N_c}$ and the chiral condensate
to $N_c$, we rescale the three-color values by factors
$\sqrt{2/3}$ and $2/3$, respectively. Namely, we choose the model
parameters such as to reproduce the values $f_\pi=75.45$ MeV,
$m_\pi=140$ MeV and $\langle\bar qq\rangle_0^{1/3}= - 218$ MeV.

Let us now turn to the choice of the regularization scheme. One
possibility is to introduce a form factor $U_\Lambda$ such that

\begin{eqnarray}
\sum_{l=0}^\infty  \int_{-\infty}^\infty \frac{dk_3}{2 \pi}
\rightarrow \sum_{l=0}^\infty  \int_{-\infty}^\infty \frac{dk_3}{2
  \pi} \ U_\Lambda\left(\sqrt{k_3^2 + 2 l|q_f| B}\right)~.
\end{eqnarray}
In fact, most of the studies of the effect of magnetic fields
within NJL-type models that include color pairing interactions are
based on this kind of regularization (see, for example,
Refs.~\cite{Fukushima:2007fc,Noronha:2007wg,Fayazbakhsh:2010gc,Mandal:2012fq}).
As for the explicit form of the regularization function one might,
in principle, be tempted to use a simple step function $\theta(x -
\Lambda)$. It is known, however, that this procedure introduces
strong unphysical oscillations in the behavior of different
quantities as functions of the magnetic field. A discussion on
this can be found in, e.g.,
Refs.~\cite{Campanelli:2009sc,Frasca:2011zn,Gatto:2012sp}, where it
was also observed that the use of smooth regulator functions
improve the situation. In fact, this allows to identify possible
physical oscillations appearing in some
cases~\cite{Fukushima:2007fc,Noronha:2007wg}. Different smooth form
factors have been used in the literature. {}For example, in
Ref.~\cite{Frasca:2011zn} Lorentzian functions of the form

\begin{equation}
U_\Lambda^{(LorN)}(x) = \left[ 1 +
  \left(\frac{x^2}{\Lambda^2}\right)^{N}\right]^{-1}~,
\label{LorN}
\end{equation}
have been used.  Alternatively, in Ref.~\cite{Fayazbakhsh:2010gc},
Wood-Saxon type form factors like

\begin{equation}
U_\Lambda^{(WS\alpha)}(x) = \left[ 1 +
  \exp\left(\frac{x/\Lambda-1}{\alpha}\right)\right]^{-1}~,
\label{WSalpha}
\end{equation}
were introduced.
Note that the form-factor in the above expressions is chosen in such a way that
that three-momentum cutoff is used concomitant with the introduction of
the magnetic field. In particular, note  that the limit of $B \to 0$
reproduces the usual three-momentum cutoff in the simpler case of a step function
regularization. But due to the aforementioned problem with this regularization, it
has motivated these other forms of regularization in the literature.
It should also be noted that all these form factors include a constant
that regulates its smoothness. To choose the values for such
constant one is limited by the fact that a too steep function
leads to the unphysical oscillations already mentioned, while a
too smooth function leads to values of the quark condensate in the
absence of the magnetic field that are quite above the
phenomenological range. Thus, the value $N=5$ is usually chosen in
the case of the Lorenztian form factor, while $\alpha = 0.05$ is
taken for the case of Wood-Saxon one.  In the following, we will
identify these two regularizations cases as Lor5 and WS0.05,
respectively.

An alternative way to regularize the finite $B$ thermodynamic
potential was suggested in Ref.~\cite{Allen15}. Within this
procedure, the terms in the thermodynamic potential that are
explicitly dependent on the magnetic field turn out to be finite
and, thus, only those terms that are independent of it need to be
regularized. In a way, this so-called ``magnetic field independent
regularization" (MFIR) can be considered as an extension of the
method described in, e.g., Ref.~\cite{Menezes:2008qt} to the case
in which color pairing interactions are present.

{}Following the regularization procedure described in
Ref.~\cite{Allen15}, after summing and subtracting convenient
terms, the thermodynamic potential Eq.~(\ref{Omega0B}) can be cast
into the form (see also App.~\ref{subtr} for details)

\begin{eqnarray}
\Omega _{0}\left( m,\Delta,B,\mu\right)  & = & \Omega_0 -
\frac{N_{c}}{2\pi ^{2}}\sum_{f = u}^{d}\left( \left|
q_{f}\right|B\right)^{2} \left\{\zeta^{\prime}\left(-1,x_{f}\right) -
\frac{1}{2}\left(x_{f}^{2} - x_{f}\right) \ln \left(x_{f}\right) +
\frac{x_{f}^{2}}{4}\right\}   \nonumber \\ & - &
\frac{N_{c}}{4\pi^{2}}\sum_{f = u}^{d}\left(\left|q_{f}\right|B\right)
\int_{0}^{\infty }dk_{3}\left\{\sum_{l = 0}^{\infty} \alpha_l
F\left(k_{3}^{2} + 2l\left|q_{f}\right|B\right) - 2
\int_{0}^{\infty}dy~F\left(k_{3}^{2} + 2y \left|q_{f}\right| B
\right)\right\}\,,
\label{ThPotential}
\end{eqnarray}
where

\begin{equation}
F\left(z^{2}\right) = \sum_{s=\pm1} \left[ \sqrt{\left(\sqrt{z^{2} + m^{2}} +
  s \mu\right)^{2} + \Delta^{2}} -  \sqrt{z^{2} + m^{2} + \Delta^{2}}\right]\,,
\label{Fz}
\end{equation}
and $x_{f}=(m^{2}+\Delta^{2})/(2\left| q_{f}\right| B)$.  In this
expression, a three-dimensional sharp cutoff $\Lambda$ can be
introduced to regularize $\Omega_0$, which has no explicit magnetic
field
dependence, while the other terms are ultraviolet finite. The
details concerning the derivation of Eq.~(\ref{ThPotential}) can
be found in Ref.~\cite{Allen15}. {}For completeness, the main ones
are also given in the App.~\ref{subtr}.

{}For each of the regularization schemes mentioned above, Lor5,
WS0.05, as well as for the MFIR case, we need to evaluate the
parameters that reproduce the scaled physical values of $f_\pi,
m_\pi$ and $\langle\bar qq\rangle_0$,  in the $N_c = 2$ case. The
procedure to obtain these quantities is discussed in detail in
Ref.~\cite{reports}. The values obtained and used in this work are
shown in Tab.~\ref{pnjl}.

\begin{table}[ht]
\caption{\label{pnjl} Parameter sets we have used for the two-color NJL
SU$_f$(2) model. Here, $m(0)$ stands for the dressed quark mass at
$\mu=B=0$.}
\begin{center}
\begin{tabular}{ccccc}
  \hline
\hspace*{.2cm} Parameter set \hspace*{.2cm} & \hspace*{.2cm} $m(0)$ (MeV)
\hspace*{.2cm}& \hspace*{.2cm} $m_c$ (MeV) \hspace*{.2cm} & \hspace*{.2cm}
$ G $ (GeV$^{-2}$)\hspace*{.2cm}    & \hspace*{.2cm} $\Lambda$ (MeV)
\hspace*{.2cm}
\\ \hline
Lor5    &    337.232 &  5.398     &  8.00             &   616 \\
WS0.05  &    311.865 &  5.401     &  7.39             &   650 \\
MFIR    &    305.385 &  5.400     &  7.23             &   657 \\ \hline
\end{tabular}
\end{center}
\end{table}

\section{Numerical Results}
\label{results}

In the previous section we gave the mean-field thermodynamic
potential in the presence of an external  magnetic field,
Eq.~(\ref{Omega0B}), and presented some different regularization
schemes to avoid the divergences. Given the corresponding
regularized form $\Omega_0^{reg}(m,\Delta,B,\mu)$, the associated
gap equations for $m$ and $\Delta$ then read
\begin{equation}
 \frac{\partial\Omega^{reg}_0(m,\Delta,B,\mu)}{\partial m} = 0\,, \;\;\;
 \frac{\partial\Omega^{reg}_0(m,\Delta,B,\mu)}{\partial \Delta} = 0\,.
 \label{gaps}
\end{equation}
In what follows we present and discuss the results obtained by
solving numerically these equations~\footnote{In our notation $m_\pi = 0.14$ GeV
and $m_\pi(eB)$ is a function of magnetic field.}.

\subsection{Behavior of the order parameters}

\begin{figure}[!htb]
  \vspace{0.6cm} \centering
  \includegraphics[width=0.48\linewidth]{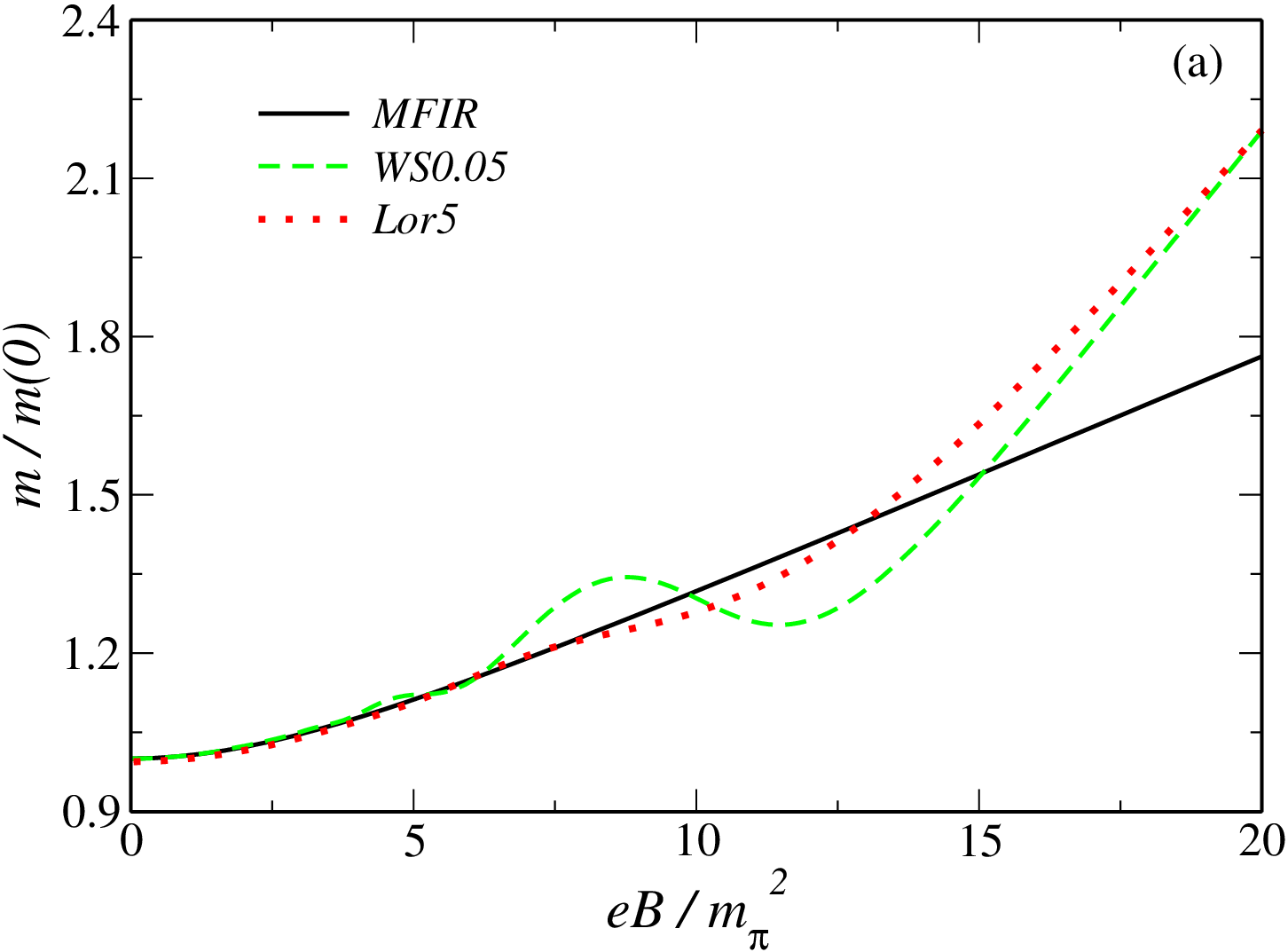}\hspace{0.1cm}
  \includegraphics[width=0.48\linewidth]{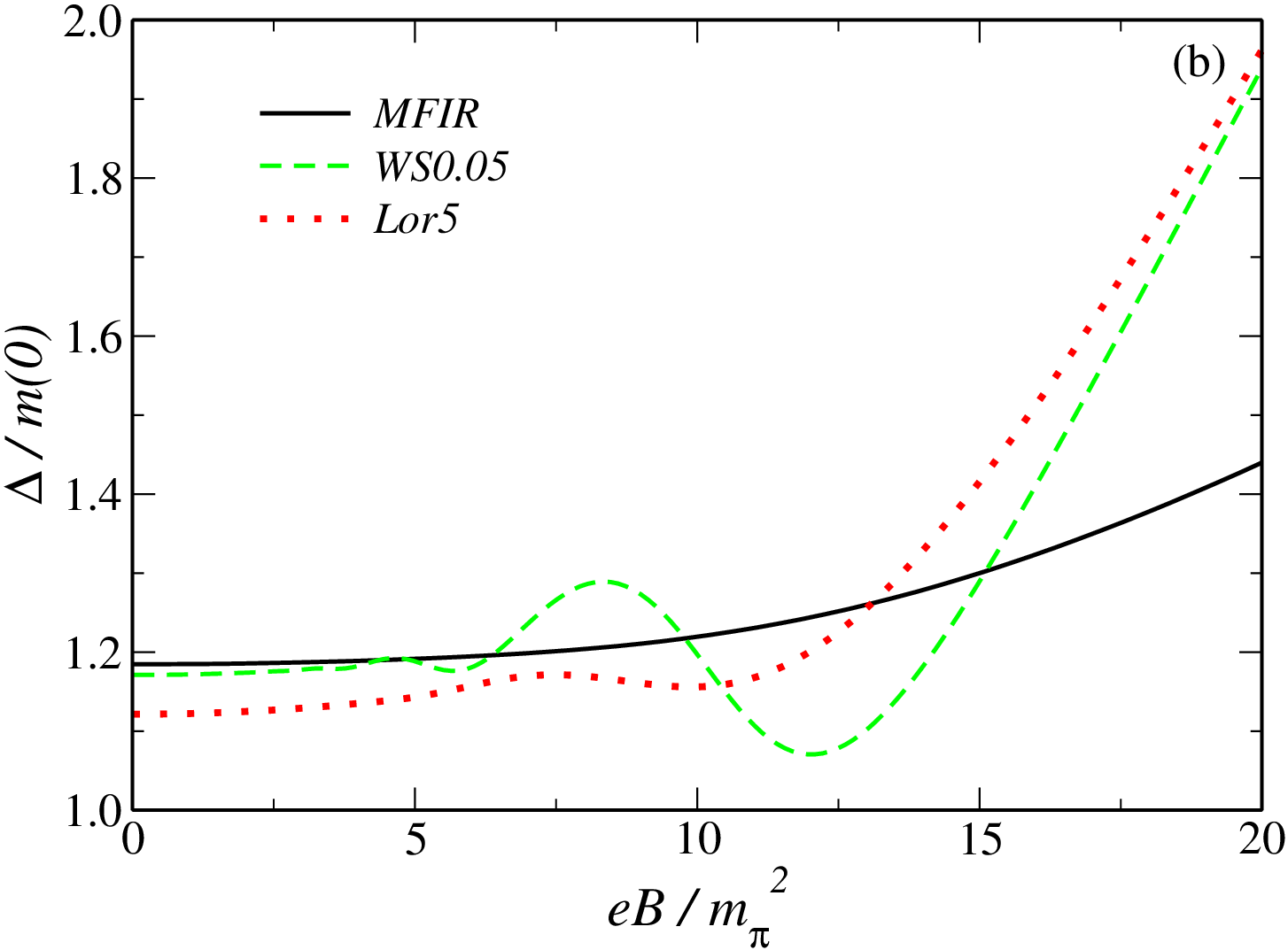}
 \caption{Results for: (a) The effective quark mass $m$, at $\mu=0$;
   and (b) for the diquark condensate $\Delta$, at
$\mu_B=6~m_\pi$,  as functions  of the magnetic field for
   the three regularization schemes (see Tab.~\ref{pnjl}).  Both $m$
   and $\Delta$ are scaled by $m(0) \equiv m(\mu = eB = 0)$.
\label{fig1}}

 \end{figure}

It is known that at zero chemical potential the magnetic catalysis
effect is present (see, for example, Ref.~\cite{Shovkovy:2012zn}).
This is seen in {}Fig.~\ref{fig1}(a), where the behavior of $m$ as
a function of the magnetic field is shown, for all three
regularization schemes studied in this work. This coincides with
the result obtained in the NJL model without the diquark
interaction since $\Delta$ vanishes in this case. For $B=0$, a
transition to a $\Delta \neq 0$ phase occurs at
$\mu=m_\pi/2$~\cite{Nishida:2005ds,Sun:2007fc}, or, equivalently, at $\mu_B =
m_\pi$. Here, $\mu_B = 2 \mu$ is the baryon chemical potential
for the $N_c=2$ case we are dealing with. Thus, fixing the
chemical potential at a value $\mu_B > m_\pi$ a non-vanishing
value for $\Delta$ is obtained. The corresponding behavior as a
function of the magnetic field is shown in {}Fig.~\ref{fig1}(b)
for the case for a value of $\mu_B=6.0\, m_\pi (\equiv 0.84\,{\rm GeV})$
(which is chosen as a representative value for illustrative purposes) and the
three regularizations schemes, WS0.05, Lor5 and MFIR.

The aforementioned oscillations are clearly seen in both panels in
{}Fig.~\ref{fig1}, i.e., in both the effective quark mass and the
diquark condensate, when using the WS0.05 and Lor5 regularization
schemes. We note that they are somewhat smaller in the second
case, which corresponds to a smoother regulator. These
non-physical oscillations can be traced to the fact that the
regularization procedure depends explicitly on the magnetic field.
Therefore, they disappear when the divergent terms are
disentangled from the magnetic contributions by using the MFIR
scheme. It is also clear from both panels in {}Fig.~\ref{fig1}
that the WS0.05 and Lor5 regularization schemes start to deviate
strongly from the MFIR scheme at large values of the magnetic
field, $eB/m_\pi^2 \gtrsim 15$. With respect to this, it should be
borne in mind that some estimates~\cite{Ferrer:2010wz} indicate
that the magnetic fields at the center of magnetars can be as
large as $e B/m_\pi^2 \simeq 30$.

\begin{figure}[htb]
  \vspace{0.6cm} \centering
  \includegraphics[width=0.48\linewidth]{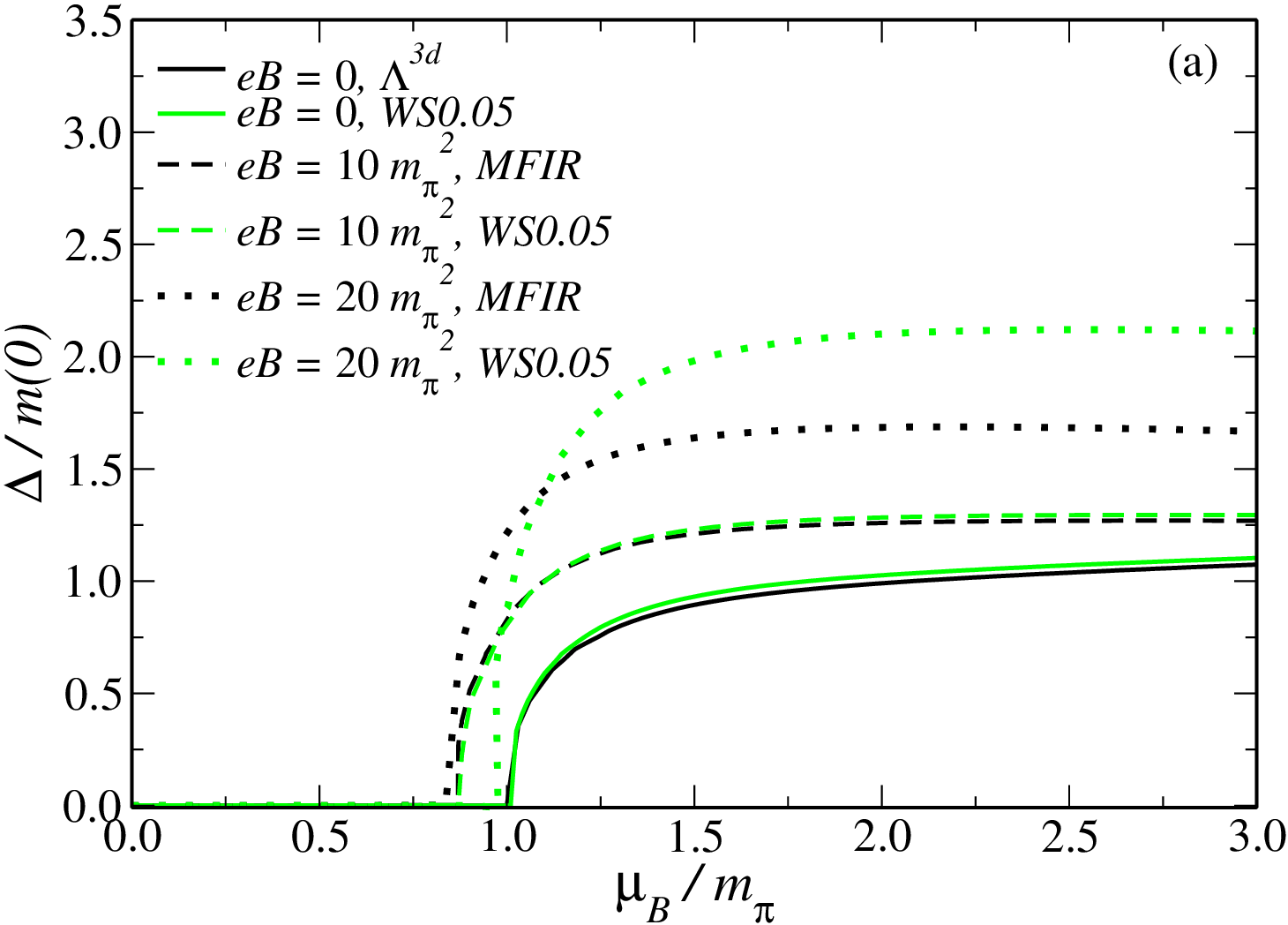}\hspace{0.1cm}
  \includegraphics[width=0.48\linewidth]{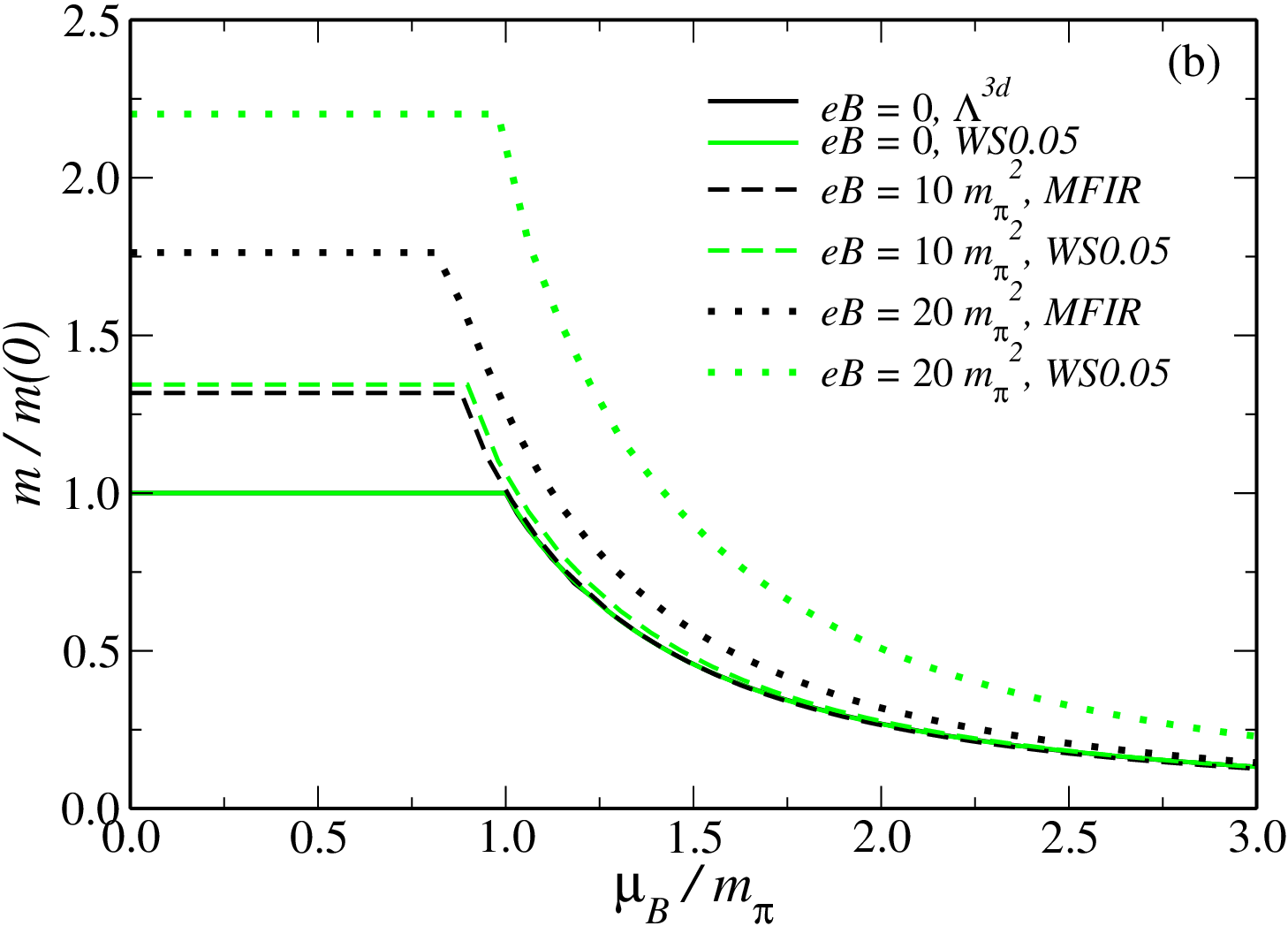}\\
 \caption{(Color online) Results for: (a) the diquark condensate $\Delta$; and (b)
   for the effective quark mass $m$, both as a function of the
   baryon chemical potential, $\mu_B$  (where $\mu_B = N_c \,\mu$ with $N_c=2$), in
   the WS0.05 and MFIR schemes.}
  \label{m-DeltaXmuB}
 \end{figure}

We consider now the behaviors of $\Delta$ and the effective quark
mass as functions of the baryon chemical potential for different
values of the magnetic field. This is shown in
{}Fig.~\ref{m-DeltaXmuB}. We restrict ourselves to the results for
the MFIR and WS0.05 regularizations, since the results for Lor5
are qualitatively similar to those for the WS0.05. There is a chirally
broken phase with $\Delta=0$ for low enough
chemical potentials and a BEC phase in which $\Delta$ is finite
and chiral symmetry is partially restored. The transition
connecting these two phases is of second-order and, when $B=0$, it
occurs at $\mu_B = m_{\pi}$. This result is in agreement with
those obtained using chiral effective
theories~\cite{Kogut:2001na}. This result also remains true for both
schemes presented here. The magnetic catalysis effect can also be
observed, where both $m$ and $\Delta$ are seen to increase with the
magnetic field. We also note that for $B=0$ (corresponding to the
lowest curves in both panels of {}Fig.~\ref{m-DeltaXmuB}), the
MFIR regularization exactly reduces to the traditional
three-dimensional cutoff $\Lambda$, indicated by the
$\Lambda^{3d}$ label in the figures, and that the results obtained
for both regularization methods are similar for $eB =0$ and $10\
m_\pi^2$, but differ significantly for $eB = 20\ m_\pi^2$.

\subsection{The critical chemical potentials}

We now determine how the magnetic field affects the critical
baryon chemical potential $\mu_{B_c}^{BEC}$ for the diquark condensate
(BEC) phase transition and  for the BEC-BCS crossover $\mu_{B_c}^{BEC-BCS}$.

Since the phase transition for diquark condensation remains
second-order  even in the presence of an external magnetic field,
we can obtain the corresponding critical baryon chemical
potential by using a Ginzburg-Landau expansion for the
thermodynamic potential Eq.~(\ref{Omega0B}) around the critical point~\cite{He:2010nb},
\begin{equation}
 \Omega_0(m,\Delta,B,\mu_B/2) = \Omega_0(m,0,B,\mu_B/2) + \alpha_2(m,B,\mu_B/2) |\Delta|^2
 + \alpha_4(m,B,\mu_B/2) |\Delta|^4 + \mathcal{O}(|\Delta|^6)~,
\label{landau}
\end{equation}
where the Ginzburg-Landau coefficients $\alpha_n$ are defined as
\begin{eqnarray}
 \alpha_n(m,B,\mu) & = & \frac{1}{n!}
 \left.\frac{d^n\Omega_0(m,\Delta,B,\mu_B/2)}{d\Delta^n}\right|_{\Delta =
   0}~.
 \label{alphan}
\end{eqnarray}
As customary,
the critical chemical potential for the BEC transition can be
obtained from the condition $\alpha_2(m,B,\mu^{BEC}_{B_c}/2) = 0$, where $m$ results from
minimizing the first term in Eq.~(\ref{landau}) for each value of $eB$.

We now turn to the determination of the critical chemical
potential for the BEC-BCS crossover. {}For this purpose, it is
convenient to define a reference chemical potential, $\mu_N =
\mu_B/2 - m$, in terms of which the transition between the two
states is characterized by the condition
$\mu_N=0$~\cite{Sun:2007fc}. The rationale behind this is as
follows. Considering for simplicity the case $B = 0$, a typical
dispersion relation is given by

\begin{equation}
 E_{\Delta}^{\pm} = \sqrt{\left(\sqrt{\vec{k}^2 + m^2} \pm
   \frac{\mu_B}{2}\right)^2 + \Delta^2}~,
\end{equation}
where $E_{\Delta}^{-}$ corresponds to particle and
$E_{\Delta}^{+}$ corresponds to antiparticle excitations. {}For
small $\mu_B$ we have $\mu_N < 0$, and the minimum of the
dispersion is located at $|\vec{k}| = 0$, with particle gap energy
$\sqrt{\mu_N^2 +  \Delta^2}$. This corresponds to the fermionic
(quark) spectrum in the BEC state. On the other hand, at larger
values of $\mu_B$, we have $\mu_N > 0$. The minimum of the
dispersion is shifted to  $|\vec{k}| = \mu_B/2$ and the particle
gap is $\Delta$. This corresponds to the fermionic spectrum in the
BCS state. Thus, the condition $\mu_N=0$ can be used to determine
the position of the crossover. Note that it implies $m =
\mu_{B_c}^{BEC-BCS}/2$. Therefore, the critical chemical potential can be
obtained by  setting this relation in the gap
equations~(\ref{gaps}) and by solving them for $\Delta$ and
$\mu_{B_c}^{BEC-BCS}$ for different values of $B$.

\begin{figure}[htb]
  \vspace{0.6cm} \centering
  \includegraphics[width=0.48\linewidth]{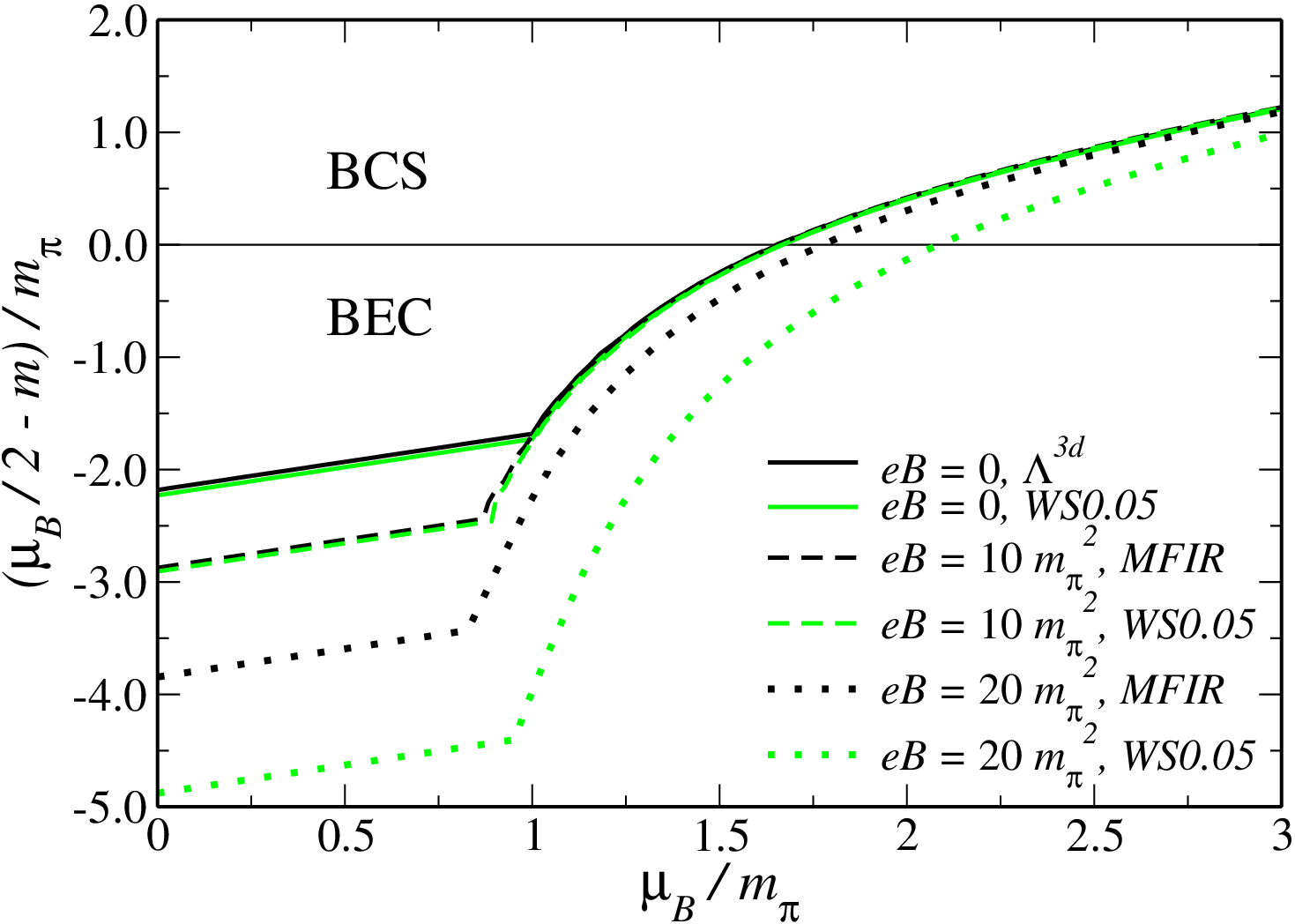}\\
 \caption{(Color online) The reference chemical potential, $\mu_N = \mu_B/2 - m$, as a
   function of the chemical potential in the WS0.05 and MFIR
   regularization schemes.}
  \label{muN-comp}
 \end{figure}

In {}Fig.~\ref{muN-comp} we show the reference chemical potential
$\mu_N$ as a function of the chemical potential, for $eB = 0,\, 10\,
m_{\pi}^2$ and $20 \,m_{\pi}^2$. We have restricted ourselves again,
for simplicity, to show results only for the MFIR and WS0.05
regularization schemes. Negative values of $\mu_N$ correspond to
the BEC region, while positive values correspond to the BCS
region. These two phases are indicated by the regions below and
above, respectively, the thin horizontal solid line shown in
{}Fig.~\ref{muN-comp}.  {}For the values of magnetic fields  shown
in {}Fig.~\ref{muN-comp}, we find that the point for which $\mu_N$
changes sign, the BEC-BCS crossover, is shifted to the right,
i.e.,  when increasing the value of the magnetic field, the
crossover takes place at a larger value of $\mu$ (or, analogously,
at a larger value of $\mu_B$). This corresponds to a strengthening
of the BEC region. This phenomena is observed to occur in both
regularization schemes considered in this work.

Note that the magnetic field values considered in the figures, $eB
= 10\ m_{\pi}^2$ and $20\ m_{\pi}^2$ (corresponding to
approximately 0.2 GeV$^2$ and 0.4 GeV$^2$ respectively) are
usually referred to in the literature as strong magnetic fields.
These values are representative of the fields expected in
heavy-ion collisions experiments, like in the RHIC and in the LHC,
which are estimated to have magnitudes of the order of  $eB \sim
15\ m_{\pi}^2$~\cite{Skokov:2013}.

\begin{figure}[htb]
  \vspace{0.6cm} \centering
  \includegraphics[width=0.48\linewidth]{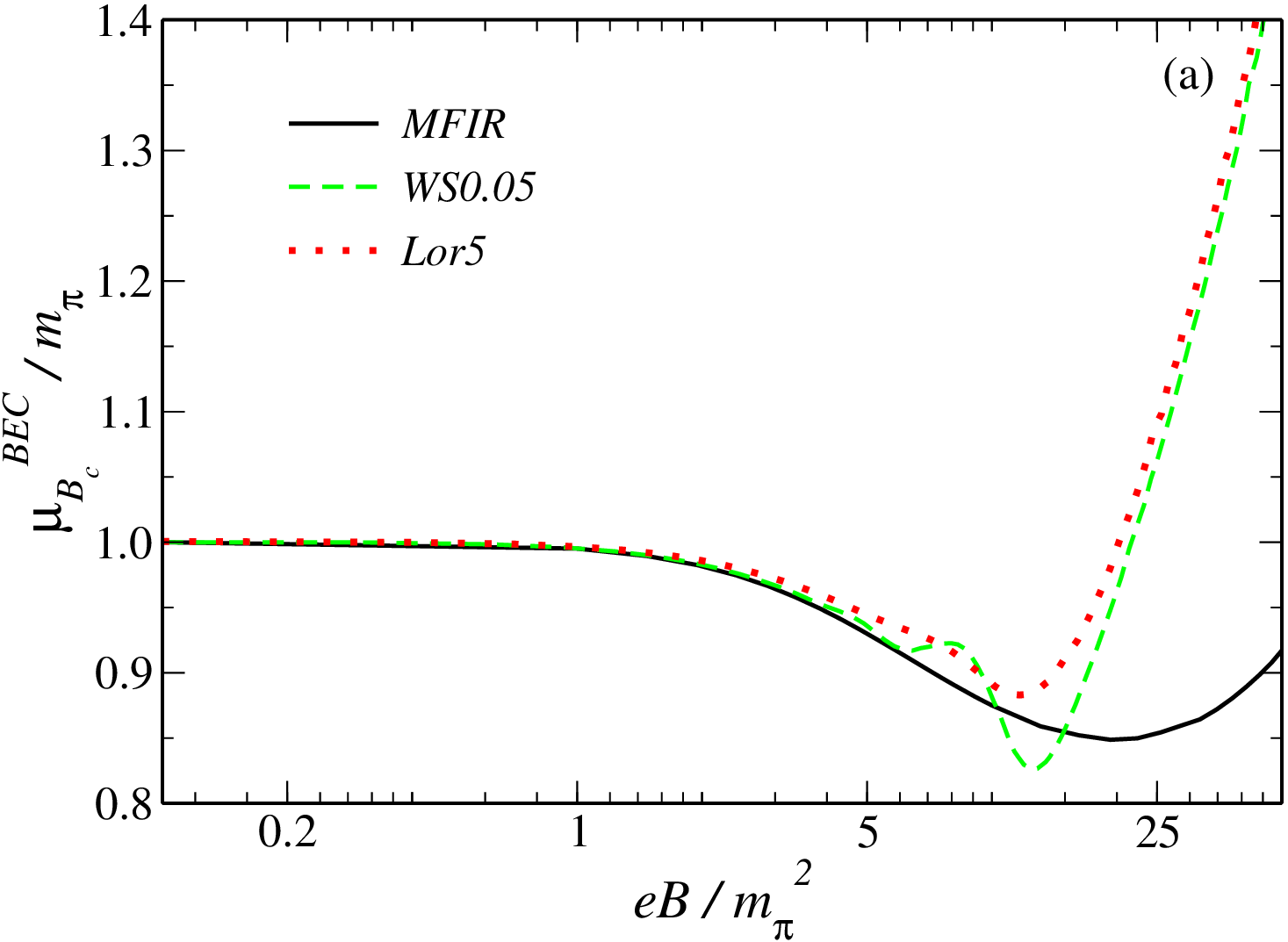}\hspace{0.1cm}
  \includegraphics[width=0.48\linewidth]{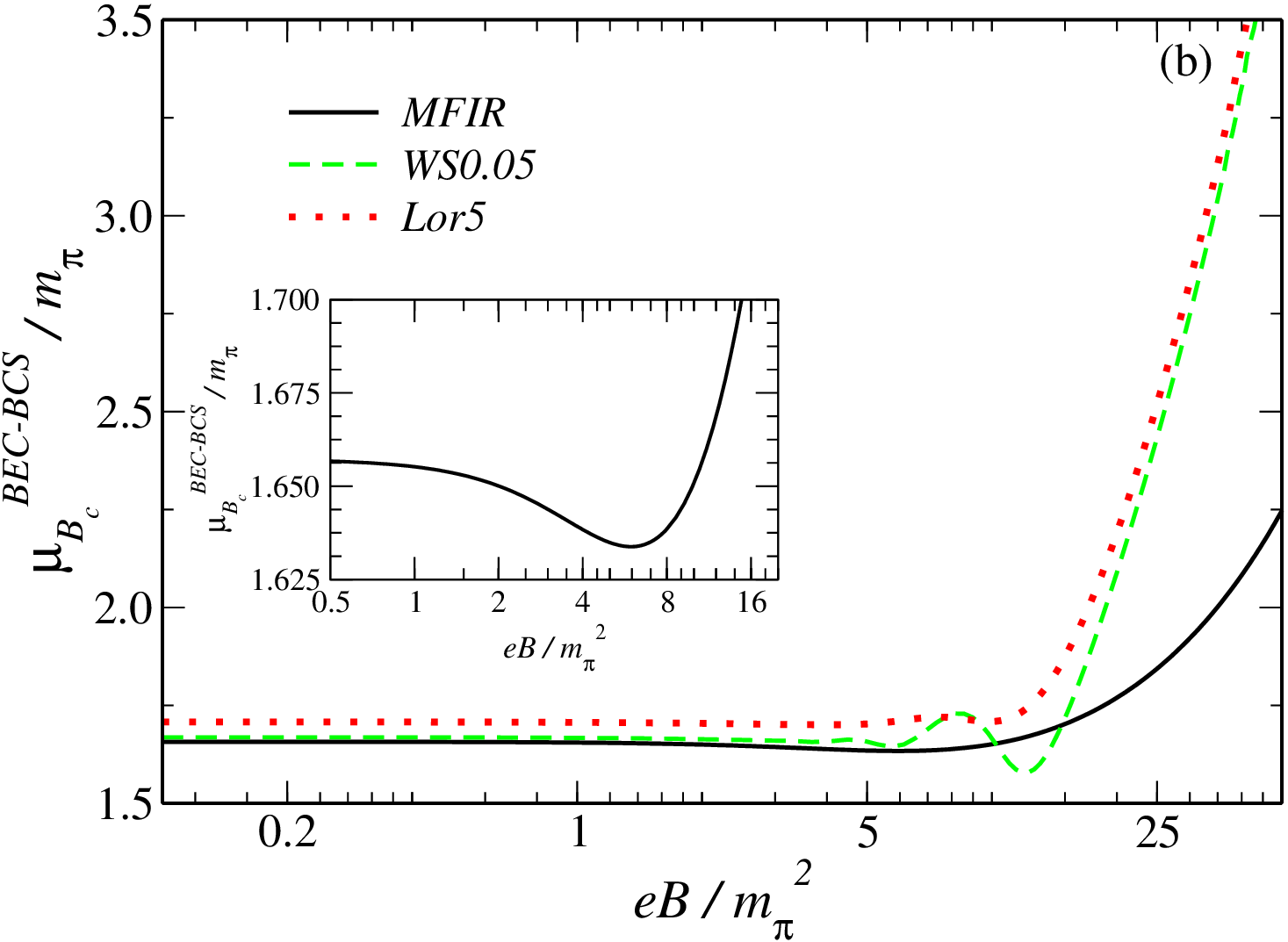}
 \caption{Critical baryon chemical potential $\mu_B$ as a function
   of the magnetic field for (a) the diquark BEC phase transition and
   (b) for the BEC-BCS crossover, in the WS0.05, Lor5 and MFIR regularization
   schemes. The inset in (b) shows the region for IMC for the MFIR case.}
  \label{muC}
 \end{figure}

The numerical results for the baryon critical chemical potential
for the BEC transition $\mu^{BEC}_{B_c}$ are shown in {}Fig.~\ref{muC}(a).  In the
MFIR scheme, we see that from small to moderately strong values of
$eB$ an inverse magnetic catalysis (IMC) effect is observed, i.e.,
the value of $\mu^{BEC}_{B_c}$ decreases with the magnetic field, while
for very strong magnetic fields, $\mu^{BEC}_{B_c}$ grows rapidly with
$eB$ in the MFIR scheme. The IMC phenomenon was first observed in
the NJL model in  Ref.~\cite{IMCklimenko} at $T = 0$ and in
Ref.~\cite{IMC2} for the full  $T-\mu-B$ case. Since then, it has
been found and described in different models (see, e.g.,
Refs.~\cite{andreas,Kneur:2013cva} and references therein). {}For
the parameters we have considered, the IMC effect is clearly seen
in the interval $1 \lesssim eB/m_\pi^2 \lesssim 20$.

The behavior of
$\mu^{BEC-BCS}_{B_c}$ for the BEC-BCS crossover is shown in
{}Fig.~\ref{muC}(b). Note that from small to moderate magnetic
field values, $ eB/m_\pi^2 \lesssim 9$, the value  of $\mu^{BEC-BCS}_{B_c}$
changes little. This indicates that the magnetic field does not
tend to affect strongly the size of the BEC region in that range of
values for the magnetic field.
Yet, we can still see an IMC effect happening also for the
crossover, which is shown in the inset in {}Fig.~\ref{muC}(b).
However, this phenomenon happens over a much smaller range of
magnetic fields and the variation of the critical baryon
chemical potential is much smaller than the one in the BEC
transition. On the other hand, for larger values of the magnetic
fields, the BEC region increases with the magnetic field, as
indicated by the rise of the critical chemical potential in
{}Fig.~\ref{muC}(b). The reduced IMC effect in the BEC-BCS crossover can
be qualitatively understood as a competition between magnetic catalysis, which
always tends to increase the chiral mass $m$ (see
{}Fig.~\ref{fig1}), thus pushing the critical chemical potential
up, and the IMC, which tends to facilitate the BEC transition.
The overall result is an almost constant critical chemical
potential from small to moderate values of the magnetic field and
an increasing $\mu^{BEC-BCS}_{B_c}$ afterwards.

{}Finally, note also the differences in behaviors for the
different regularization schemes seen in {}Fig.~\ref{muC}. Both
WS0.05 and Lor5 regularization schemes produce again non-physical
oscillations similar to what we have seen before in
{}Fig.~\ref{fig1}. It is important to remark that, for the chosen
parametrizations, there are no oscillations related to the van
Alphen-de Haas (vA-dH) effect in this model. These oscillations do
have a physical meaning and would usually be expected to appear as
a consequence of the oscillations of the density of states near
the Fermi surface. In the present case they are not visible
because $\Delta$ takes rather large values in the crossover side,
while in the BEC side the effective quark mass remains also rather
large. As a consequence, the oscillations are washed away in both
cases. Studies have been reported in which the oscillations are
actually visible, owing to a smaller diquark coupling
\cite{Noronha:2007wg,Fukushima:2007fc} or because there are quark
species that do not participate in the pairing (as would occur in
the 2SC phase for color SU(3)), which have ordinary vA-dH
oscillations, which in turn produce oscillations indirectly on
$\Delta$ \cite{Allen15}. In the later case, the quarks that are
decoupled from the diquark gap produce vA-dH transitions, which
are related to the change in population of Landau levels, and are
nearly vertical in the phase diagram for chemical potential and
magnetic field. Note also that the WS0.05 and Lor5 regularization
schemes always overestimate the values for the critical baryon chemical potentials, both for
the BEC second-order diquark condensate phase transition and also
for the BEC-BCS crossover, with the difference increasing with the
magnetic field. The reason for this is the same as already
explained before with respect to the same behavior seen in
{}Fig.~\ref{fig1}.

\section{Interpreting the results}
\label{results2}

Let us better interpret the numerical results shown in the previous
section. {}For the numerical results presented in this section, we have restricted to
show only for the MFIR scheme for simplicity.

{}First, let us show that the condition used to determine the
BEC transition point remains valid. The transition point was determined
by setting the coefficient for the quadratic term in the diquark field in
Eq.~(\ref{landau}) to zero, i.e., by making $\alpha_2(m,B,\mu^{BEC}_{B_c}/2) = 0$.
This condition is, of course, only valid if the transition is second order.
We can verify that the transition is in fact second order and
remains so in a finite magnetic field by analyzing  the coefficient for the quartic term
in the diquark field in Eq.~(\ref{landau}). If the quartic term is positive for
the values of magnetic field considered, i.e., $\alpha_4(m,B,\mu^{BEC}_{B_c}/2) > 0$
across the BEC transition, then the transition remains second order.

\begin{figure}[htb]
  \vspace{0.6cm} \centering
\includegraphics[width=0.48\linewidth]{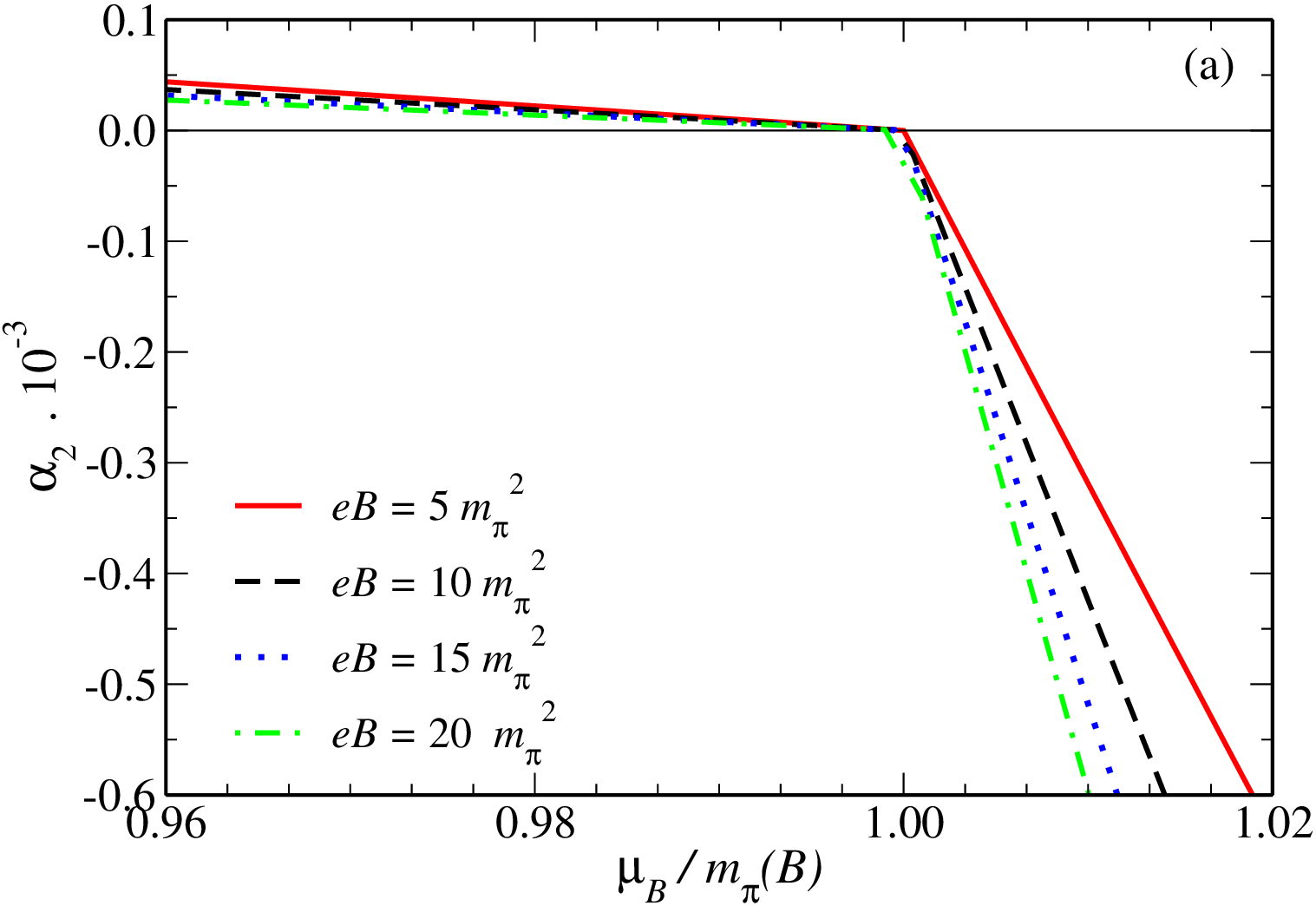}
\includegraphics[width=0.48\linewidth]{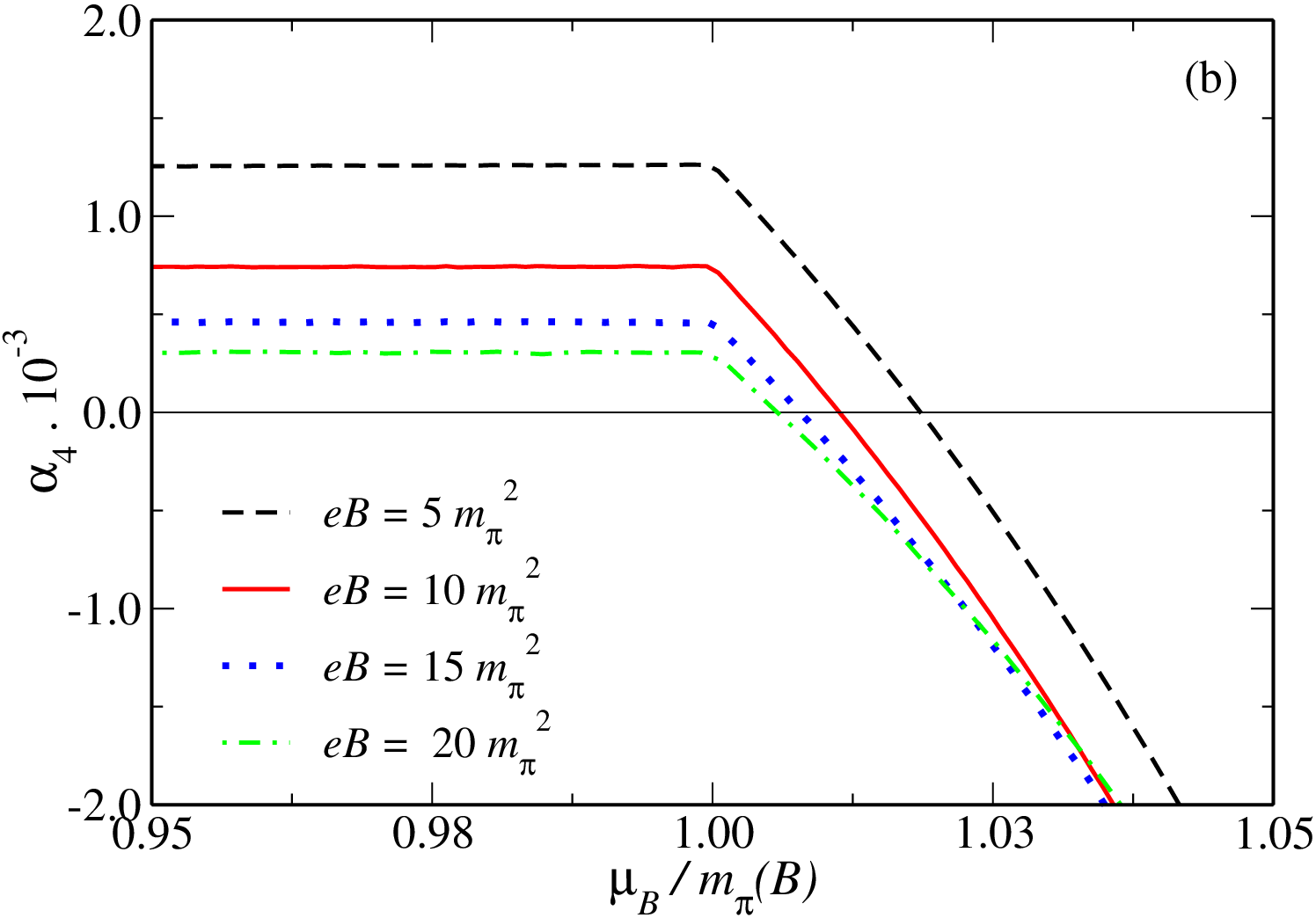}
 \caption{The quadratic and quartic coefficients in the Ginzburg-Landau expansion for the
thermodynamic potential, Eq.~(\ref{landau}).}
  \label{alpha4plot}
 \end{figure}

In {}Fig.~\ref{alpha4plot} we plot both $\alpha_2(m,B,\mu_{B}/2)$ and $\alpha_4(m,B,\mu_{B}/2)$ for a reasonable
range of chemical potentials and for different values for the magnetic field.
Note that in the horizontal axis in {}Fig.~\ref{alpha4plot} we are using the baryon chemical potential
normalized by the pion mass computed at a given value of magnetic field,  $\mu_B/m_\pi(B)$. The reason for
presenting the results this way will be
fully justified below, where we explicitly give the expression defining the pion mass in a finite
magnetic field, $m_\pi(B)$. As well-known, at $B=0$ the BEC transition for diquark condensation happens at
$\mu_B = m_\pi$. As shown in the previous section, in the presence of a magnetic field the BEC transition
tends to happen for a different value for $\mu_B$. The value for $\mu_B$ where the transition happens
can be either smaller, in the case of IMC, or higher, as in the case where magnetic catalysis (MC) becomes
dominant, than in the case where $B=0$. However, the BEC transition can be shown to always happen
at the value of the pion mass where it is now determined at the given value of the magnetic field, $\mu_B=m_\pi(B)$.
The plot in {}Fig.~\ref{alpha4plot}(a) indeed confirms this.
It shows that $\alpha_2(m,B,\mu_{B}/2)= 0$ at  $\mu_B=m_\pi(B)$
for all values of magnetic field.

{}Finally, in {}Fig.~\ref{alpha4plot}(b), we also show that $\alpha_4(m,B,\mu_{B}/2)$
remains positive at the point where $\mu_B=m_\pi(B)$ and, therefore, the BEC transition
remains second-order at finite magnetic fields. We also notice from the results
in {}Fig.~\ref{alpha4plot} that $\alpha_4$ becomes negative for values of $\mu_B$ slight higher
than the BEC transition point. But this does not indicate that a first-order phase transition
would follow the BEC transition, since this happens where $\alpha_2$ is already negative.
Hence, no additional (local) vacuum values for the diquark field can be generated.

The dependence of the critical chemical potential for the BEC
phase can be related to the magnetic field-dependence of the
diquark mass $m_d$ (at zero chemical potential). As the diquark is
an electrically neutral particle, its mass should be insensitive
to $B$ to leading order, and it will mildly decrease as $B$ gets
bigger. In the $N_c=2$ case we are dealing with, at $\mu=0$ the
diquark is degenerated with $\pi_0$ and we can analyze what
happens with $m_d$ looking at the neutral pion
mass~\cite{pionpol}. As an aside observation concerning this
degeneracy between the scalar diquarks with the neutral pion, we
can interpret it based on some symmetry arguments satisfied by the
model. Note that in the absence of magnetic field the symmetry
group of the present NJL model is $Sp(4)$ (see, e.g.,
Ref.~\cite{ChPT} for associated analysis in two color QCD) with
scalar diquarks and pions lying in the corresponding 5-dim irrep.
This symmetry, however, gets broken
in the presence of the magnetic field, since the charged particles acquire a
different mass than the neutral ones. Thus, the 5 dim irrep goes to a 3-dim 
one for the neutral sector, containing the two diquarks and the neutral pion.

The neutral pion mass decreases in real QCD, and it justify
our IMC results shown in {}Fig.~\ref{muC}(a) (at least below $10 \,m_\pi^2$). A similar situation
is seen to happen in real QCD with isospin chemical potentials (see, e.g., Ref.~\cite{endrodi14}),
where the BEC phase also contains charged pions. Accordingly, the magnetic field enhances the
critical chemical potential, since the charged pion mass increases with $B$.

To further understand the IMC seen in the numerical results shown in
{}Fig.~\ref{muC}, we can first establish a relation between the critical baryon
chemical potential and the effective mass for the (neutral) pion field.
Note that in the absence of the magnetic field the BEC transition happens at
$\mu^{BEC}_{B_c} = m_\pi$, as is well known and explicitly seen from the results of the
previous section. As we have also shown above, at a finite magnetic field this
transition also remains of second order. It is natural to then expect that
the condition for the transition, $\mu^{BEC}_{B_c} = m_\pi$, remains valid also at finite magnetic
field values if we replace the vacuum pion mass by its effective value at
some magnetic field, i.e., $m_\pi(B)$. Since the BEC transition is for diquark condensation
and the diquarks are neutral here, we can try to relate the transition point
directly in terms of the neutral pion effective mass.

The effective pion mass is determined from its inverse propagator, which in momentum
space-time can be expressed as

\begin{equation}
G_{\pi}(p_0,\vec p) = -\frac{1}{2 G} - \Pi_{\pi}(p_0,\vec p)\,,
\label{Gpi}
\end{equation}
where $\Pi_{\pi}(p_0,\vec p)$ is the self-energy contribution for the
pion, coming from a quark-antiquark loop at
leading one-loop order. Since we are interested in the pion mass,
we evaluate the self-energy in the rest frame, $\vec p=0$ and then set
the energy at the mass-shell, $p_0 = m_\pi$, where $m_\pi$ is to be interpreted
as the effective pion mass. The effective pion mass is then determined by the
pion pole condition, $G_{\pi}(p_0=m_\pi,\vec p=0) =0$, or

\begin{equation}
\frac{1}{2 G} = - \Pi_{\pi}(p_0=m_\pi,\vec p=0)\,.
\label{pionpole}
\end{equation}

In the rest frame and computed at zero temperature,
the one-loop self-energy for the pion in the present two-color NJL model
can be expressed as~\cite{Strodthoff:2011tz}

\begin{eqnarray}
\Pi_{\pi}(p_0=m_\pi,\vec p=0) = 4N_c \int \frac{d^3k}{(2 \pi)^3}
\frac{E_{\Delta}^{+}E_{\Delta}^{-} + \epsilon_k^+ \epsilon_k^- +\Delta^2}
{m_\pi^2 - (E_{\Delta}^{+} + E_{\Delta}^{-} )^2 }
\left( \frac{1}{E_{\Delta}^{+}} + \frac{1}{E_{\Delta}^{-}} \right)\,,
\label{Pi}
\end{eqnarray}
where we have defined $E_{\Delta}^{\pm}= \sqrt{ (\epsilon_k^\pm)^2 + \Delta^2}$,
$\epsilon_k^\pm = E_k \pm \mu$ and $E_k = \sqrt{k^2 + m^2}$, as also defined earlier in
Sec.~\ref{Thpot}.  At the BEC transition point for
diquark condensation, we can set $\Delta =0$ in Eq.~(\ref{Pi}) and the
pion pole condition becomes

\begin{equation}
\frac{1}{2 G} =  16N_c \int \frac{d^3k}{(2 \pi)^3}
\frac{E_k}{4 E_k^2-m_\pi^2 }\,.
\label{pionpoleBEC}
\end{equation}
To introduce the magnetic field, we can follow the prescription as given in
Eq.~(\ref{Btrick}), to then obtain that

\begin{equation}
\frac{1}{2 G} =  8N_c \sum_{f
    = u}^{d}\frac{\left| q_{f}\right| B}{4\pi} \sum_{l = 0}^{\infty }
  \alpha_{l} \int_{-\infty}^{+\infty} \frac{dk_{3}}{ 2\pi }
\frac{ E_{k_3,l}}{4  E_{k_3,l}^2-m_\pi^2 }\,.
\label{BpionpoleBEC}
\end{equation}

An explicit derivation of the pion one-loop polarization term at zero temperature and in an external magnetic
field can be found in Ref.~\cite{pionpol}. In that work the authors evaluated the masses of
$\sigma$ and $\pi_0$, the $\pi^0$ decay constant in the presence of a magnetic field at vanishing
temperatures and baryon densities. In particular, they show that the Gell-Mann-Oakes-Renner (GOR) relation remains valid
in a magnetic medium.

We can now compare the pion pole condition Eq.~(\ref{BpionpoleBEC}) with the
one determining the BEC transition point, which is determined by
setting the coefficient of the quadratic term in the diquark field in
Eq.~(\ref{landau}) to zero, $\alpha_2(m,B,\mu^{BEC}_{B_c}/2) = 0$. Using
Eq.~(\ref{alphan}), we obtain that

\begin{equation}
\alpha_2(m,B,\mu^{BEC}_{B_c}/2) = 0 =\frac{1}{4 G} -  \frac{N_c}{2}\sum_{f
    = u}^{d}\frac{\left| q_{f}\right| B}{4\pi} \sum_{l = 0}^{\infty }
  \alpha_{l} \int_{-\infty}^{+\infty} \frac{dk_{3}}{ 2\pi }
\left[ \frac{1}{ E_{k_3,l} + \mu^{BEC}_{B_c}/2 } + \frac{1}{ E_{k_3,l} - \mu^{BEC}_{B_c}/2 }
\right]\,.
\label{alpha2}
\end{equation}
The above equation can be rewritten in the form

\begin{equation}
\frac{1}{2 G} =  8N_c\sum_{f
    = u}^{d}\frac{\left| q_{f}\right| B}{4\pi} \sum_{l = 0}^{\infty }
  \alpha_{l} \int_{-\infty}^{+\infty} \frac{dk_{3}}{ 2\pi }
\frac{ E_{k_3,l}}{4  E_{k_3,l}^2- (\mu^{BEC}_{B_c})^2 }\,.
\label{BECcond}
\end{equation}
Comparing Eq.~(\ref{BpionpoleBEC}) with Eq.~(\ref{BECcond}) it becomes
clear that the BEC condition for diquark condensation, $\mu^{BEC}_{B_c} = m_\pi$,
also remains valid at finite magnetic field, i.e.,
$\mu^{BEC}_{B_c} = m_\pi(B)$, where $m_\pi(B)$ is the solution of Eq.~(\ref{BpionpoleBEC})
along the BEC transition.

\begin{figure}[htb]
  \vspace{0.6cm} \centering
  \includegraphics[width=0.48\linewidth]{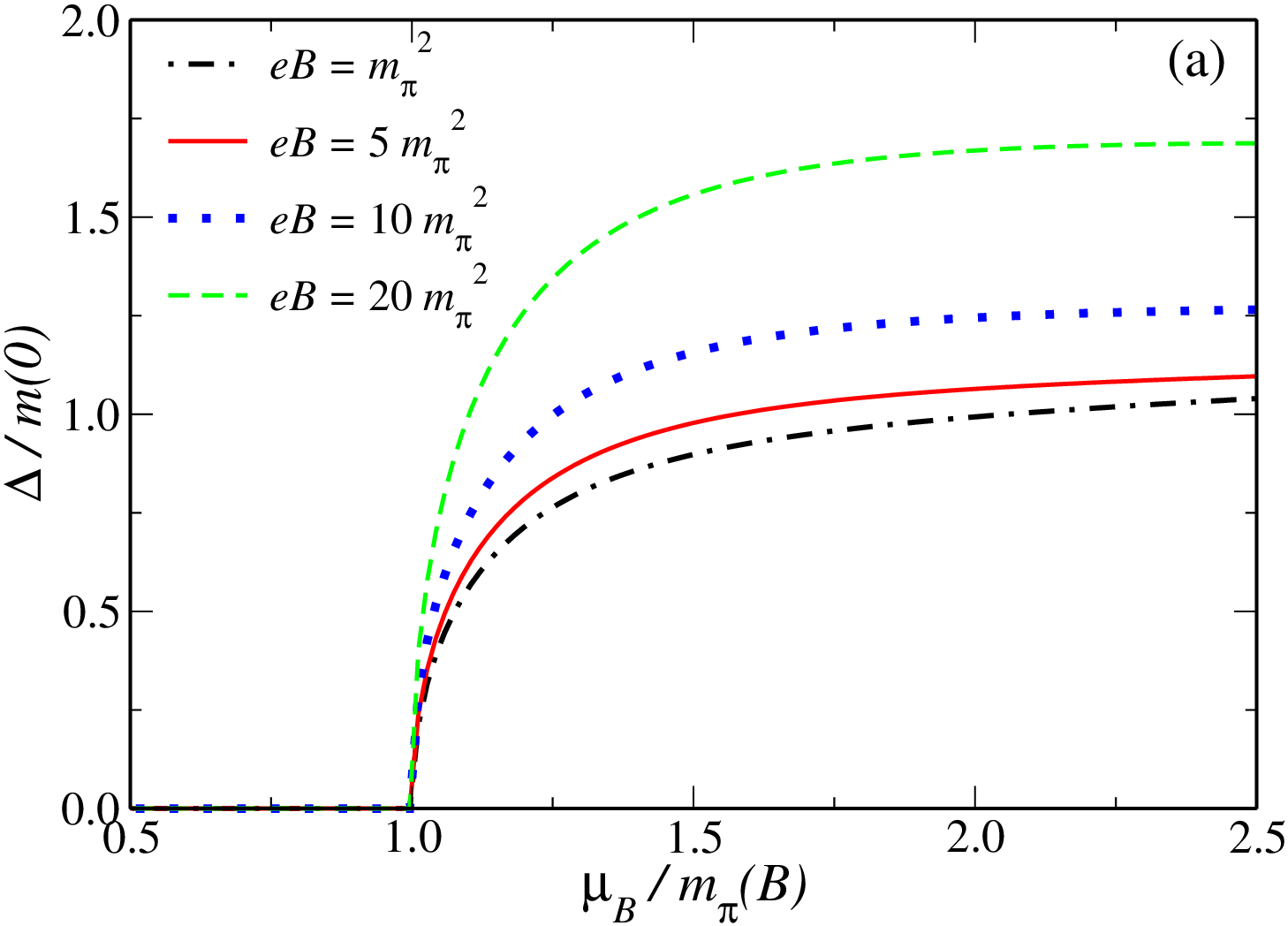}\hspace{0.1cm}
  \includegraphics[width=0.48\linewidth]{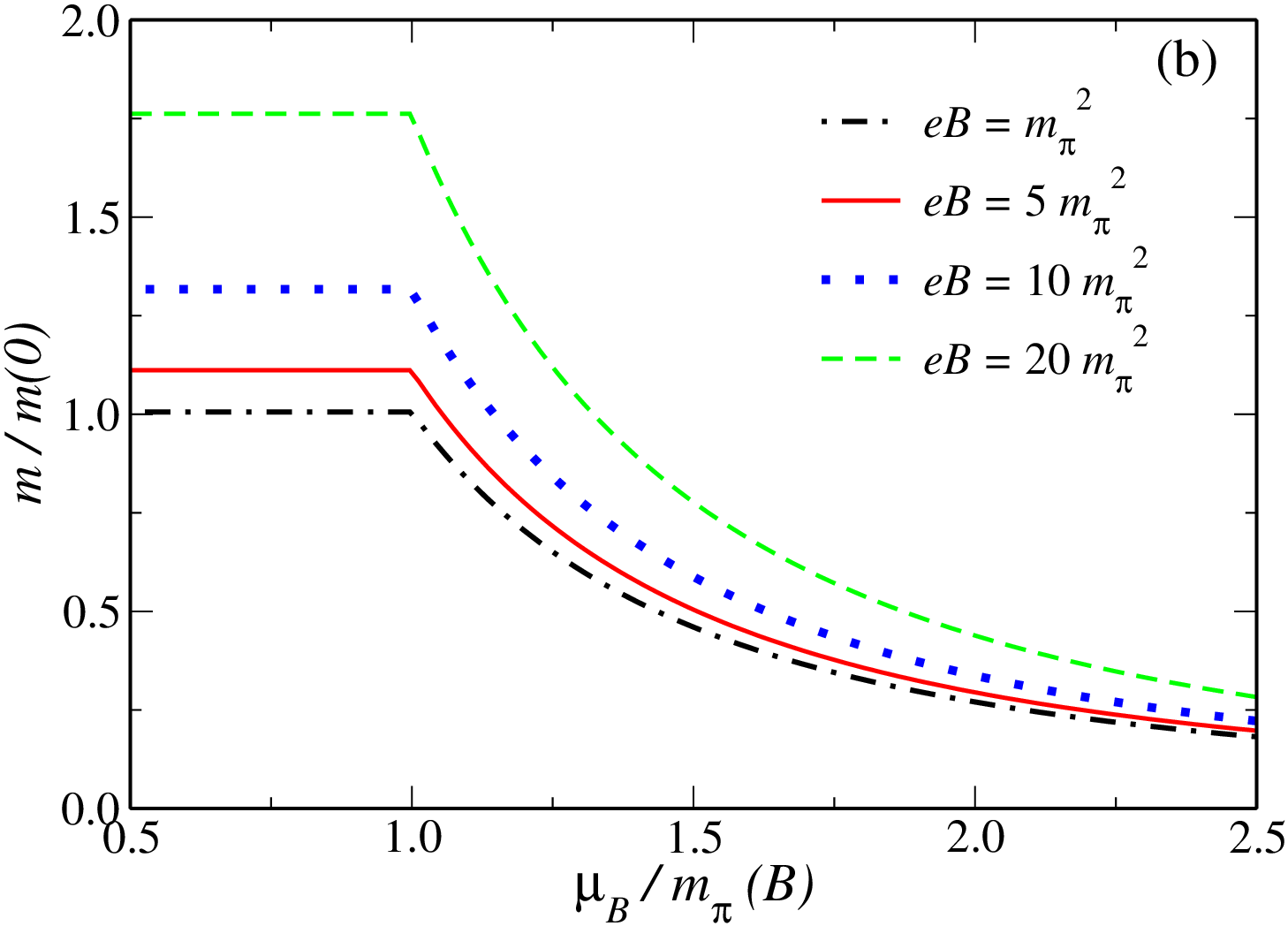}
 \caption{(a) Diquark condensate $\Delta$; and (b) effective quark mass $m$, both as a
 function of the baryon chemical potential $\mu_B$ in the MFIR
   scheme. Here the pion mass used to rescale $\mu_B$ is a function of the magnetic field
   $m_\pi(B)$.}
  \label{delta_m_muBompione}
 \end{figure}

In {}Fig.~\ref{delta_m_muBompione} we show our numerical results for the diquark condensate
$\Delta$ and for the effective quark mass $m$ as a function of $\mu_B$ for different values of the
magnetic field. Note that we have normalized $\mu_B$ with respect to $m_\pi (B)$. We can
then observe that indeed the BEC transition always happens at $\mu^{BEC}_{B_c} = m_\pi(B)$.

In Refs.~\cite{lhe_2005,lhe_review} the authors show analytically that the ``chiral rotation"
behavior predicted by Chiral Perturbation Theories (ChPT)~\cite{Ratti:2004ra,Brauner:2009gu,chiralrot}
is valid in the NJL model near the quantum phase transition. The ChPT results
for the diquark condensate and effective quark mass at $B=0$ can be found for example in Table 3 of
Ref.~\cite{ChPT}. We have verified that the relations found in the framework of ChPT~\cite{ChPT}
at vanishing magnetic field are also very well satisfied in our case at finite values of $B$.
The relevant expressions at finite $B$ are given in Table~\ref{chptB}, where $m(B,0)=m(B,\mu_B=0)$.

\begin{figure}[htb]
  \vspace{0.6cm} \centering
  \includegraphics[width=0.48\linewidth]{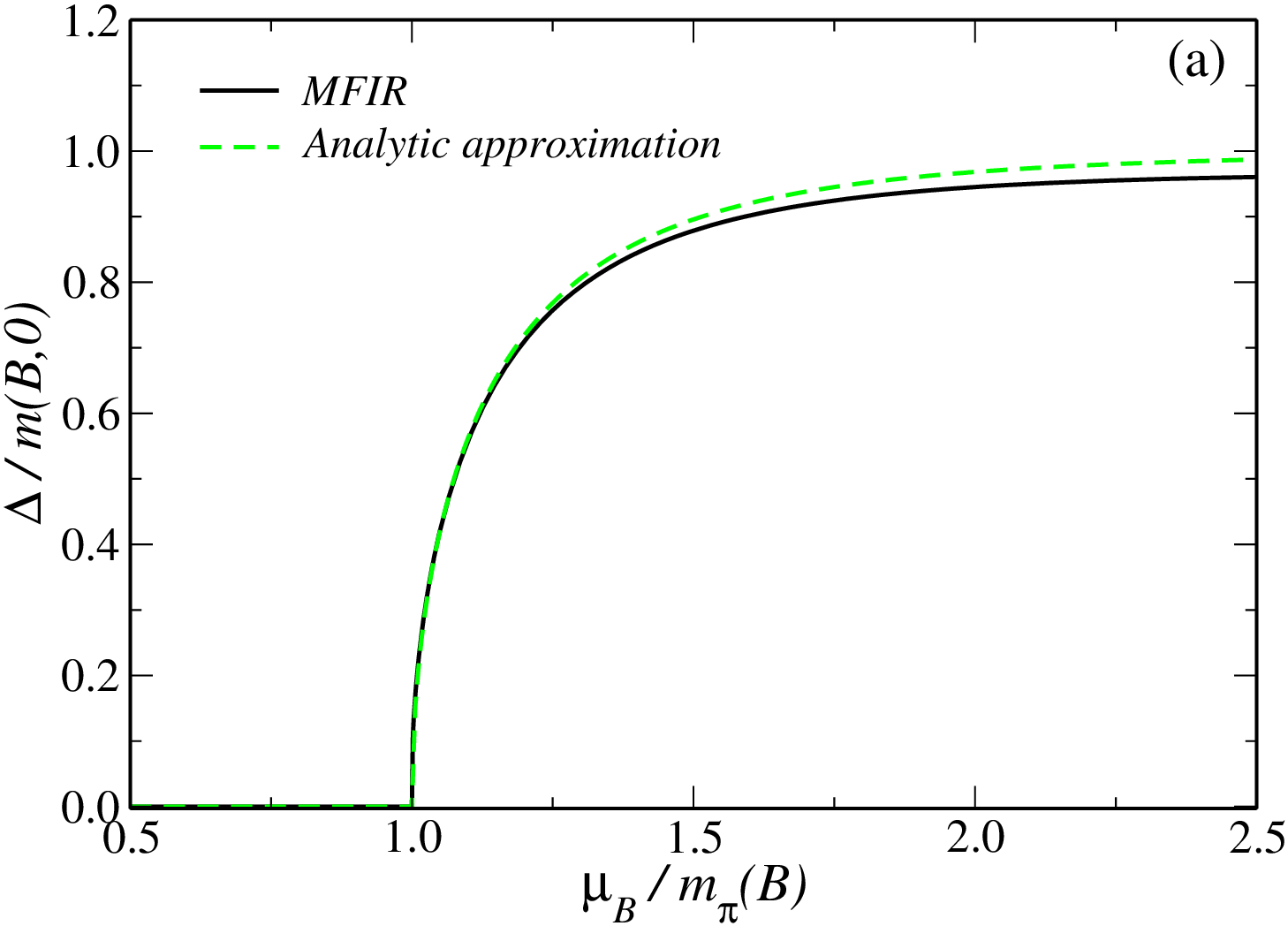}\hspace{0.1cm}
  \includegraphics[width=0.48\linewidth]{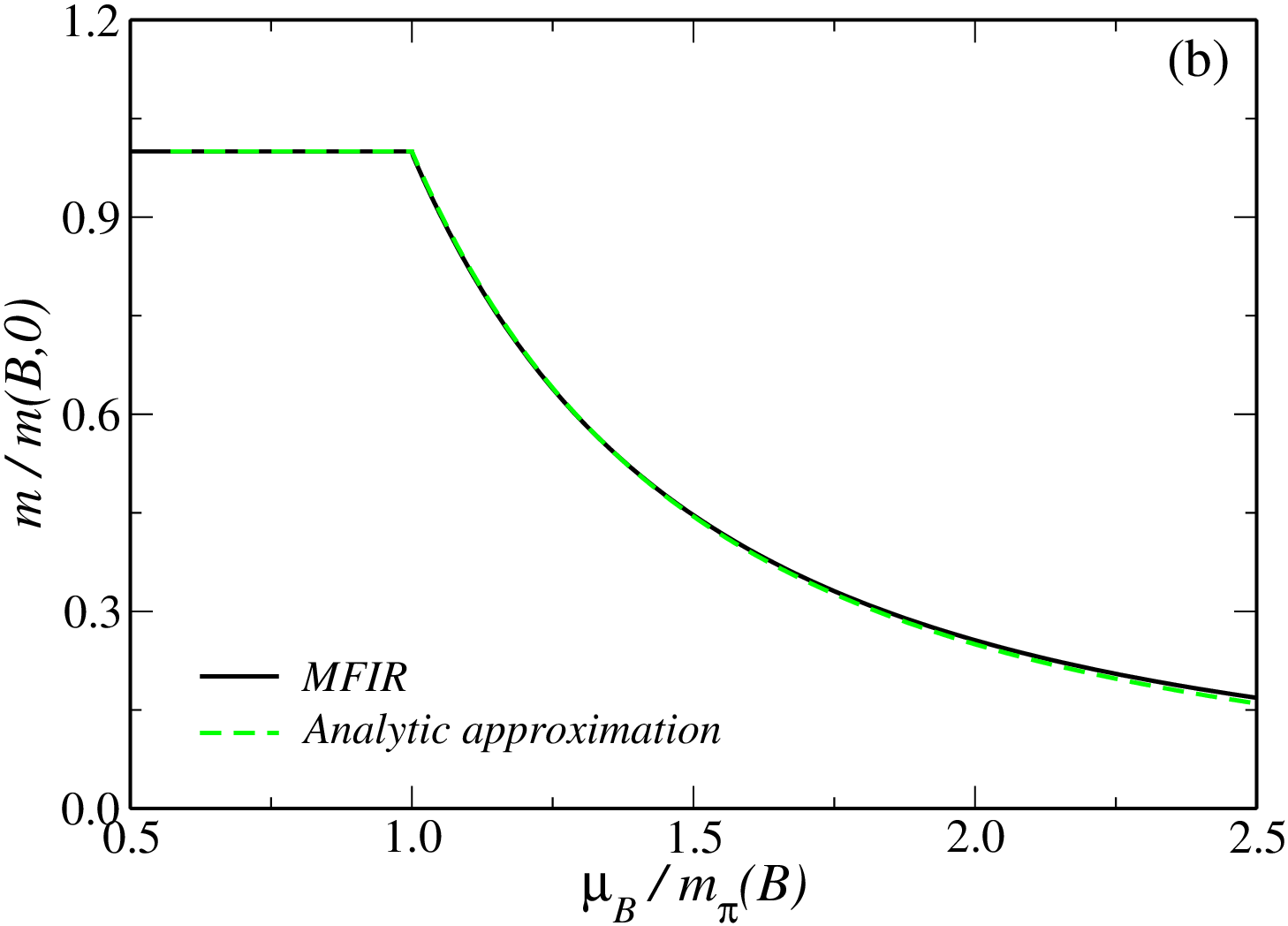}
 \caption{(a) Diquark condensate $\Delta$; and (b) effective quark mass $m$, both as a
 function of the baryon chemical potential $\mu_B$ for $eB = 10 m_\pi^2 $. Solid lines evaluated
 with  MFIR scheme and dashed lines are for our analytical approximation.}
  \label{delta_m_ChPT}
 \end{figure}

\begin{table}[h]
\caption{\label{chpt} The mass $m$ and diquark $\Delta$ in the two phases of the theory.}
\begin{center}
\begin{tabular}[c]{c| c| c}
  \hline  \hline
 phase               &  $m$                & $\Delta$    \\ \hline
$\mu_B < m_{\pi}(B)$    &    $m(B,0)$                  &  0   \\ \hline
$\mu_B \geq m_{\pi}(B)$    &    $m(B,0)\left[\dfrac{m_{\pi}(B)}{\mu_B}\right]^2$ &
$m(B,0)\sqrt{1-\left[\dfrac{m_{\pi}(B)}{\mu_B}\right]^4}$    \\
\hline
\end{tabular}
\end{center}
\label{chptB}
\end{table}
In {}Fig.~\ref{delta_m_ChPT} we numerically verify the validity of the relations given in Tab.~\ref{chptB}
at finite values of the magnetic field. The difference between the approximate relations and the numerical results
are at the percentage level.

\begin{figure}[htb]
  \vspace{0.6cm} \centering
  \includegraphics[width=0.48\linewidth]{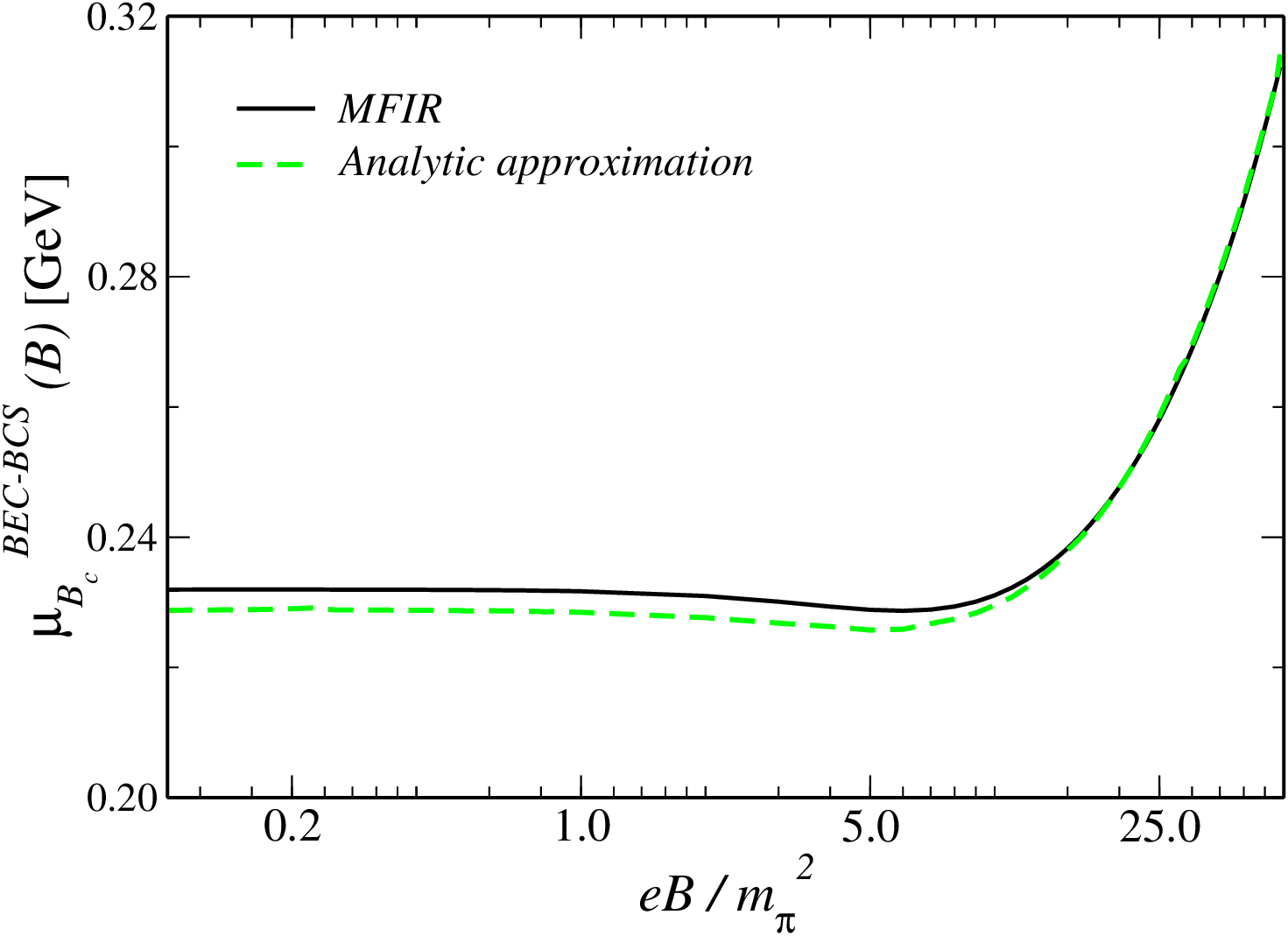}
 \caption{The analytical approximation, Eq.~(\ref{apprrelation}) (dashed line) and the
numerical result (solid line) for magnetic field dependence of the critical chemical potential
for the crossover BEC-BCS $\mu_B^{BEC-BCS}(B)$ (here expressed in units of GeV).}
   \label{appmuBcBECBCS}
 \end{figure}

Having better understood the results for the BEC transition, we can now move to the
case of the BEC-BCS crossover.
The authors of Ref.~\cite{Sun:2007fc}  give a very simple approximate analytic
expression for the position of the BEC-BCS crossover (see, for instance, Eq.~(40) in that reference),
which only depends on the constituent quark mass and on the pion mass in the vacuum,
$\mu_B^0=\left[ 2m(0)m_\pi^2 \right]^{1/3}$.
We can show in the present study that a similar expression also holds when accounting for the magnetic field dependence
of $m$ and $m_\pi$. Moreover, we have already proved analytically that
$\mu^{BEC}_B = m_{\pi}(B)$. Using this relation, the properly extended version of
Eq.~(40) given in Ref.~\cite{Sun:2007fc}, expressing both the effective quark mass and pion mass
as functions of the magnetic field,

\begin{equation}
\mu_B^{BEC/BCS}(B) \simeq \left[ 2 m(B)m_\pi^2(B) \right]^{1/3}\,,
\label{apprrelation}
\end{equation}
turns out to be valid for all the values of the magnetic considered in the present work.
In {}Fig.~\ref{appmuBcBECBCS} we show our numerical results for $\mu_B^{BEC/BCS}(B)$
and the analytical expression obtained at finite $B$. It is clear that the result
Eq.~(\ref{apprrelation}) holds in the present study.

The expression Eq.~(\ref{apprrelation})
is particularly useful to better understand the results of the previous section,
shown in {}Fig.\ref{muC}(b), as far the behavior with the magnetic field is concerned.
{}For small/intermediate values of $B$, the decrease of
$\mu_B^{BEC}(B)$ is basically compensated by the increase of the
effective quark mass that follows from the MC effect. At around $eB = 10\,m_{\pi}^2$, both
$m_{\pi}(B)$ and $m(B)$ display a sharp increase as a function of the magnetic field, leading to
the observed behavior of $\mu_B^{BEC/BCS}(B)$.

\section{Conclusions}
\label{concl}

In this work we studied the effect of the application of an
external magnetic field on the BEC-BCS crossover. The model
considered was the two-color NJL with diquark interactions.  When
including magnetic fields, special attention has to be paid to the
regularization scheme used.  We have considered in our studies
three forms of regularization: one with the Lorenztian form factor
(Lor5), another one using a Wood-Saxon form factor (WS0.05) and a
recently studied regularization procedure (MFIR) in which the
contributions that are explicitly  dependent on the magnetic field
are finite and, thus, do not require to be regularized. As shown
in Ref.~\cite{Allen15} in the two-flavor NJL model and also here
when a diquark interaction term is included, the MFIR scheme fully
prevents unphysical oscillatory behavior from appearing in the
physical quantities, like in the condensates and in the critical
chemical potentials.

Our results show that there is an IMC effect in the critical
chemical potential for the BEC transition for diquark
condensation. This is clearly observed for magnetic fields in the
range $1 \lesssim eB/m_\pi^2 \lesssim 20 $. The critical chemical
potential decreases during the IMC by at most $15\%$, favoring the
BEC transition. {}For larger values of the magnetic field, only
magnetic catalysis is observed, requiring larger chemical
potentials for the diquark condensation. With the further increase
of the chemical potential, the system reaches the BEC-BCS
crossover. The corresponding critical chemical potential is very
weakly affected by the magnetic field  up to values $eB \simeq 9\
m_\pi^2$. There is still an IMC happening also for the crossover,
but the phenomenon is much weaker than the one seen in the BEC
transition with diquark condensation. The critical chemical
potential for the crossover decreases at most by only $1.5\%$.
{}For larger magnetic fields, $eB \gtrsim 9\ m_\pi^2$, the BEC-BCS
crossover does however exhibit a magnetic catalysis effect. Thus,
strong magnetic fields tend to strengthen the BEC region, once the
difference $\mu_B /2- m$ becomes positive at a higher value of
$\mu_B$.

\appendix
\section{The MFIR scheme}
\label{subtr}

Let us briefly describe the MFIR scheme applied to the
thermodynamic potential Eq.~(\ref{Omega0B}). We follow closely the
procedure shown in Ref.~\cite{Allen15}, where the interested
reader can find more details.

Let us consider the contribution of each flavor to the second term
in Eq~(\ref{Omega0B}). It reads

\begin{equation}
I_f=\frac{\left|q_{f}\right| B}{2\pi} \sum_{s=\pm 1} \sum_{l =
0}^{\infty } \alpha_{l}\int_{-\infty }^{+\infty
}\frac{dk_{3}}{2\pi} \  \sqrt{(E_{k_3,l} + \ s \mu)^2 + \Delta^2}
~. \label{defI0}
\end{equation}
Summing and subtracting a similar expression but where $\mu$ is
set to zero, we get

\begin{eqnarray}
I_f & = & \frac{\left|q_{f}\right| B}{2\pi}\sum_{l =
  0}^{\infty}
\alpha_{l}\int_{-\infty}^{+\infty}\frac{dk_{3}}{2\pi}
F\left(k_{3}^{2} + 2l\left|q_{f}\right|B\right)
+ \frac{\left|q_{f}\right|B}{\pi} \sum_{l =
  0}^{\infty}\alpha_{l}\int_{-\infty }^{+\infty}
\frac{dk_{3}}{\left(2\pi\right)}\sqrt{E_{k_3,l}^2 + \Delta^2}~,
\label{I02}
\end{eqnarray}
where $F(z^{2})$ was defined in Eq.~(\ref{Fz}).

By adding and subtracting its form at vanishing magnetic field,
the first term on the right hand side of Eq.~(\ref{I02}) results
in

\begin{eqnarray}
\frac{\left|q_{f}\right| B}{2\pi}\sum_{l =
    0}^{\infty}
  \alpha_{l}\int_{-\infty}^{+\infty}\frac{dk_{3}}{2\pi}
  F\left(k_{3}^{2} + 2l\left|q_{f}\right|B\right) & =&
  2\int\frac{d^{3}k}{\left(2\pi\right)^{3}} F(\vec{k}^{2})
  +\frac{\left|q_{f}\right|B}{\left(2\pi\right)}
  \int_{-\infty}^{+\infty}\frac{dk_3}{\left(2\pi\right)}F\left(k_{3}^{2}\right)
  \nonumber\\ & &\!\!\!\!\!\!\!\!\!\!\!\!\!\!\!\!\!\!\!\!\!\!\!\!\!\!\!\!\!\!\!\!\!\!\!
  +\frac{2\left|q_{f}\right|B}{2\pi}
  \int_{-\infty}^{+\infty}\frac{dk_3}{2\pi}
  \left\{\sum_{l = 1}^{\infty}F\left(k_{3}^{2} +
  2l\left|q_{f}\right|B\right) -
\int_{0}^{\infty}dy~ F\left(k_{3}^{2} +  2y\left|q_{f}\right|B
\right)\right\}~, \label{T1fin}
\end{eqnarray}
where in the last term of the above expression we have used cylindrical coordinates,
$\vec k \equiv (k_\rho, k_\theta,k_3)$, performed the angular integration and
defined the new variable $y = k_{\rho}^2/(2|q_f|B)$.

The second term on the right hand side of Eq.~(\ref{I02}) can be
evaluated using the method discussed in
Ref.~\cite{Menezes:2008qt}. We get

\begin{eqnarray}
 \frac{\left|q_{f}\right|B}{\pi} \sum_{l =
  0}^{\infty}\alpha_{l}\int_{-\infty }^{+\infty}
\frac{dk_{3}}{\left(2\pi\right)}\sqrt{E_{k_3,l}^2 + \Delta^2} & =
& 4\int\frac{d^{3}k}{\left(2\pi\right)^{3}} \sqrt{k^{2} +
  m^{2} + \Delta^2}\nonumber  \\ & + &
\frac{\left(\left|q_{f}\right|B\right)^{2}}{\pi^{2}}
\left\{\zeta^{\prime}\left(-1,x_f\right) -
\frac{1}{2}\left(x_f^{2} - x_f\right) \ln\left(x_f\right) +
\frac{x_f^{2}}{4}\right\}\,,
 \label{I0Bfin}
\end{eqnarray}
where $x_f = (m^{2} + \Delta^2)/(2|q_{f}|B)$ and  $\zeta'(s,a)$ is
the $s$-derivative of the Hurwitz zeta function~\cite{zeta}.

Replacing Eqs.~(\ref{T1fin}) and (\ref{I0Bfin}) in Eq.~(\ref{I02}) and
using the definition of $F(z^2)$, Eq.~(\ref{Fz}), we get

\begin{eqnarray}
I_f  & = & 2 \sum_{s=\pm
   1}\int\frac{d^3k}{(2\pi)^3}\ \sqrt{(E_k + s \ \mu)^2 + \Delta^2}~ +
\frac{ \left(\left| q_{f}\right|B\right)^{2}}{\pi^{2}}
\left\{\zeta^{\prime}\left(-1,x_{f}\right) -
\frac{1}{2}\left(x_{f}^{2} - x_{f}\right) \ln \left(x_{f}\right) +
\frac{x_{f}^{2}}{4}\right\} \nonumber
\\ & + &
\frac{\left|q_{f}\right|B}{2\pi^{2}} \int_{0}^{\infty
}dk_{3}\left\{\sum_{l = 0}^{\infty} \alpha_l F\left(k_{3}^{2} +
2l\left|q_{f}\right|B\right) - 2
\int_{0}^{\infty}dy~F\left(k_{3}^{2} + 2y \left|q_{f}\right| B
\right)\right\}~. \label{new}
\end{eqnarray}
The first term in Eq.~(\ref{new}) is exactly the $B = 0$
contribution. It is divergent, but it can now be handled through
standard methods, e.g., using a three dimensional cutoff. At the
same time, all the remaining terms in Eq.~(\ref{new})
depending on the magnetic field are finite.
Summing over flavors and using Eq.~(\ref{Omega0}) we get
Eq.~(\ref{ThPotential}).


\acknowledgments

D.C.D. and P.H.A.M. are supported by Coordena\c{c}\~ao de
Aperfei\c{c}oamento de Pessoal de N\'{\i}vel Superior
(CAPES). R.L.S.F. and R.O.R are partially supported by research grants
from Conselho Nacional de Desenvolvimento Cient\'{\i}fico e
Tecnol\'ogico (CNPq). R.O.R. is also  partially supported by a
research grant from Funda\c{c}\~ao Carlos Chagas Filho de Amparo \`a
Pesquisa do Estado do Rio de Janeiro (FAPERJ).
P. G. A. and N.N.S. are partially supported by CONICET (Argentina) under
grant PIP 00682 and by ANPCyT (Argentina) under grant
PICT-2011-0113. R.L.S.F. acknowledge the kind hospitality of the
Center for Nuclear Research at Kent State University, where part of this
work has been done. R.L.S.F. is also grateful to G. Endr\"odi for useful
correspondence and comments.


\end{document}